\documentclass{pasj01}
\draft 
\Received{$\langle$reception date$\rangle$}
\Accepted{$\langle$acception date$\rangle$}
\Published{$\langle$publication date$\rangle$}
\begin{document}
\title{ALMA Astrometry of the Objects within 0.5 pc of Sagittarius A$^\ast$}
\author{Masato Tsuboi$^{1}$, Takahiro Tsutsumi$^2$, Atsushi Miyazaki$^3$, Ryosuke Miyawaki$^4$, and Makoto Miyoshi$^5$}
\altaffiltext{1}{Institute of Space and Astronautical Science (ISAS), Japan Aerospace Exploration Agency,\\
3-1-1 Yoshinodai, Chuo-ku, Sagamihara, Kanagawa 252-5210, Japan }
\email{tsuboi@vsop.isas.jaxa.jp}
\altaffiltext{2}{National Radio Astronomy Observatory, P. O. Box O,  Socorro, NM 87801-0387, USA}
\altaffiltext{3}{Japan Space Forum, Kanda-surugadai, Chiyoda-ku, Tokyo,101-0062, Japan}
\altaffiltext{4}{College of Arts and Sciences, J. F. Oberlin University, Machida, Tokyo 194-0294, Japan}
\altaffiltext{5}{National Astronomical Observatory of Japan, Mitaka, Tokyo 181-8588, Japan}
\maketitle
\begin{abstract}
Atacama Large Millimeter/Submillimeter Array (ALMA) is promising to be a powerful tool for precision astrometry of the area around Sagittarius A$^\ast$ (Sgr A$^\ast$) because it has the high angular resolution, high sensitivity, and wide field of view. 
We have observed the area including the Nuclear Star Cluster at 230 GHz with ALMA in October 2017. The angular resolution is $\sim0\farcs03$.  We determined the relative positions to Sgr A$^\ast$ of 65 compact objects in the area with the accuracy of $\sim0\farcs001$. 
We also analyzed the similar ALMA archival data obtained in June 2019 and determined the 64 relative positions in these objects. We derived the proper motions relative to Sgr A$^\ast$ by comparing these positions. 
The derived proper motions are roughly described with both clockwise and counterclockwise rotations around Sgr A$^\ast$.  The rotation velocities are reproduced by Kepler orbits bounded around Sgr A$^\ast$.
Moreover, the proper motions include co-moving clusters for example IRS13E and IRS13N.
The positions and proper motions are almost consistent with those by previous infrared observations. 
Therefore the observational demonstrations would prove that ALMA is a powerful tool for precision astrometry of the region.  
\end{abstract}
\section{Introduction}
The Galactic center (GC) region is the central region of our Milky Way  galaxy and located at the distance of $d\sim8$ kpc from us (e.g. \cite{Boehle}, \cite{Abuter}). 
This region harbors the GC super massive black hole (GCBH) with $M_\mathrm{GCBH}\sim4\times10^6 M_\odot$ (e.g. \cite{Ghez}, \cite{Boehle}, \cite{Abuter}).  The object corresponding to the GCBH is observed as Sagittarius A$^\ast$ (Sgr A$^\ast$) in radio, infrared, and X-ray wavelengths.  In the GC region, we can find peculiar phenomena by present telescopes, which should be found in the central regions of external spiral galaxies by the future huge telescopes.
Accordingly, the GC region is considered to be a laboratory to study the central regions of spiral galaxies. Especially, the proximity makes it possible to depict the structures smaller than $1000~$au using Atacama Large Millimeter/sub-millimeter Array (ALMA). Therefore the high-angular resolution radio observations of the GC region are important as it serves as a template for studies of central regions of other spiral galaxies.

About $100$ OB and Wolf-Rayet (WR) stars have been found within $r<0.5$ pc of Sgr A$^\ast$ by high-angular resolution infrared (IR) observations in these three decades (e.g. \cite{Genzel},  \cite{Figer1999},  \cite{Figer2002}). These stars are called the Nuclear star cluster (NSC)  as a group. 
The distribution and kinematics of the stars in the NSC would provide important informations for understanding the formation mechanism of the stars, moreover for exploring the existence of the GCBH (e.g. \cite{Ghez}, \cite{Genzel2010}, \cite{Boehle}, \cite{Abuter}). 
Such studies have been performed only by sensitive infrared telescopes (IRTs) (e.g. \cite{Ott}, \cite{Maillard}, \cite{Paumard}, \cite{Schodel2009}, \cite{Muzic2008}, \cite{Fritz2010}, \cite{Eckart2013}). 
Meanwhile ALMA is promising to be also a suitable tool for precision astrometry because of the following properties. 
ALMA has a very sharpe synthesized beam with very high sensitivity, for example, $\sim0\farcs030$ with $1\sigma\lesssim50 \mu$Jy at 230 GHz (\cite{Tsuboi2019b}). This means that ALMA can observe WR stars themselves not surrounding ionized gas envelops (Cf. \cite{Zhao2009}).
The position of the stars can be measured directly by referencing the position of Sgr A$^\ast$ by ALMA because ALMA can always detect Sgr A$^\ast$.   
This is in contrast to the IRTs, which have been detecting Sgr A$^\ast$ only when its flaring phase although the latest near infrared instrument, GRAVITY on VLTI can detect Sgr A$^\ast$ on a regular basis (e.g. \cite{Boehle}, \cite{Abuter}). 
Moreover, ALMA has a wide field of view (FOV) (e.g. $\sim26\arcsec$ at 230 GHz, $\sim60\arcsec$ at 90 GHz), which correspond to the areas within the radiuses of $r\sim0.5$ pc and $r\gtrsim1$ pc of Sgr A$^\ast$. 
Therefore, the observational demonstrations would prove that ALMA is excellent for precision astrometry of the GC region. 

Throughout this paper, we adopt $d=8.0$ kpc as the distance to the Galactic center: $1\arcsec$ corresponds to 0.039 pc or 8000 au at the distance. And we use the International Celestial Reference System (ICRS) as the coordinate system. 

\section{Observation and Data Reduction} \label{sec:ODR}
\subsection{Observation in Oct. 2017}
We have observed Sgr A$^\ast$ itself and the surrounding area in the ALMA Cy.5 program (2017.1.00503.S. PI M.Tsuboi).  
The FOV is centered at  $\alpha_{\rm ICRS}$ = $17^{\rm h}45^{\rm m}40^{\rm s}.04$  and $\delta_{\rm ICRS}$= $-29^{\circ}00'28\farcs2$, which is a nominal center position of  Sgr A$^\ast$. 
The center frequencies of the spectral windows are  $f_c=217.5, 219.5, 234.0,$ and $231.9$ GHz, respectively. The frequency coverage and frequency resolution of the spectral window are 2.0 GHz and 15.625 MHz, respectively. The spectral windows except for $f_c=231.9$ GHz are used for the continuum observation.  
 The diameter of the FOV at 226 GHz is $D_{\rm FOV}\sim26\arcsec$ in FWHM, which corresponds to the physical diameter of $d\sim1.0$ pc at the distance to the Galactic center. The observations  were performed in ten days within the duration from 5 Oct. 2017 to 20 Oct. 2017. The observation epochs were in the period of the longest baseline antenna configuration of ALMA. 
J1744-3116 was used as a phase calibrator in the  226 GHz observation. The flux density scale was determined using J1924-2914. 

We have used the standard observation procedure of ALMA, which had been prepared by the Joint ALMA Observatory (JAO). After the calibrations by observing the standard sources (J1924-2914, J1752-2956, and J1744-3116), we observed the target, Sgr A$^\ast$, and phase calibrator, J1744-3116, alternately. The integration times of  Sgr A$^\ast$ and J1744-3116 in one scan were 55 s and 19 s, respectively. The duration of one scan was about 90 s. The total elapsed time of one day observation was about 5000 s. 

The flux density of Sgr A$^\ast$ were significantly changed between the first seven days (from 5 Oct. 2017 to 14 Oct. 2017) and the last three days (from 17 Oct. 2017 to 20 Oct. 2017). Accordingly, we used the first seven days of the data. The observation epoch is $t_\mathrm{obs}=2017.773$ year.
However, some intensity change of Sgr A$^\ast$ is still found even in the first seven days  (see \cite{Iwata}).
Because such intensity change is contrary to the basic assumption of synthesis imaging that sources seen in the FOV are not variable,  it makes specific artifacts in the resultant map (e.g.  \cite{Cotton}, \cite{Rau}).
One of them is a series of concentric circles centered on Sgr A$^\ast$ (see Figure 17-6 in \cite{Cotton}). When the observed source is on the artificial concentric circle,  the position and intensity may shift from the original values besides the degradation of SN. 
Because we identify that the image quality degradation by intensity variation of the major source in the FOV  is significant for synthesized imaging and astrometry, we would report this issue in the other paper.

We performed the data analysis by Common Astronomy Software Applications (CASA 5.6) \citep{McMullin}. We used the ``self-calibration" method  to obtain a high dynamic range in the continuum map. The imaging to obtain the maps was done using {\tt gaincal}, {\tt applycal}, and {\tt tclean} tasks sequentially. 
The first imaging uses only {\tt tclean} task.
In the second step,  {\tt gaincal} makes the correction table, ``gaintable", using the model from the first imaging and {\tt applycal} applies it to the visibility data. Then the second map is imaged by {\tt tclean} task using the corrected visibility data. In the next step, the second map is used to make the ``gaintable".   In the first 2 self-calibration iterations, only the phase correction was applied using the solution interval of `solint=inf' in {\tt gaincal}, which means that the solution interval is the length of a scan.
In the successive iterations, we used the solution interval of `solint=int', which means that the solution interval is the length of data sampling. Several iterations were need to get the good solution because Sgr A$^\ast$ and surrounding structures are very complicated. Finally, we applied both the phase and amplitude calibrations.
After 7 iterations,  side lobe features were disappeared even around weak objects with peak intensity of $< 1$ mJy beam$^{-1}$. Then we concluded that the imaging reached to the goal. The angular resolutions using natural weight and Briggs weight (robust parameter = 0.5)  as {\it u-v} sampling are $\theta_{{\mathrm maj}}\times\theta_{{\mathrm min}}=0\farcs037 \times 0\farcs025, PA=85^\circ$ and $\theta_{{\mathrm maj}}\times\theta_{{\mathrm min}}=0\farcs029 \times 0\farcs020, PA=86^\circ$, respectively. These correspond to the physical extents at the Galactic center distance of $d_{{\mathrm maj}}\times d_{{\mathrm min}}=0.0015$ pc$ \times 0.0010$ pc$, PA=85^\circ$ and $d_{{\mathrm maj}}\times d_{{\mathrm min}}=0.0012$ pc$ \times 0.0008$ pc$, PA=86^\circ$, respectively. The maximum detectable angular scale of this observation is approximately 10 times larger than the angular resolution, or $\sim0\farcs4$. 
It is possible that the extended emission around Sgr A$^\ast$ degrades the accuracy. Because strong extended emission, except for a few objects, is not detected brighter than the artifacts mentioned above in the image, the degradation is almost negligible.
The sensitivities using natural weight and Briggs weight (robust parameter = 0.5) in the emission-free areas are $1\sigma=12~ \mu$Jy beam$^{-1}$ or 0.36 K  in $T_\mathrm{B}$ and  $1\sigma=15~ \mu$Jy beam$^{-1}$ or 0.45 K  in $T_\mathrm{B}$, respectively. 
Although some image quality degradation mentioned above is still identified around Sgr A$^\ast$ even after applying the ``self-calibration" method, the dynamic range\footnote{This dynamic range is a nominal value. Because the Sgr A region is very crowding, it is difficult to obtain the single definite value.} reaches to $\frac{\mathrm{Flux~density~of~Sgr A^\ast}}{\mathrm{image~noise~level}} = \frac{\sim3 \mathrm{Jy}}{12~\mathrm{or}~15 \mu\mathrm{Jy}} \gtrsim200000$ using both of these weight schemes. 
The observation parameters are summarized in Table 1.

The spectral window of $f_c=231.9$ GHz is for the H30$\alpha$ recombination line ($\nu_{rest}$= 231.9009 GHz) observation. The study of gas motions using this data has been published partly in the previous paper (\cite{Tsuboi2019}).
Before imaging in the recombination line, the continuum emissions of Sgr A$^\ast$ and the GCMS were subtracted from the combined data observed in the first seven days using CASA task, {\tt uvcontsub (fitorder=1)}.  The imaging to obtain channel maps was done using CASA 5.6 with {\tt tclean} task. 
The angular resolution is $\theta_{{\mathrm maj}}\times\theta_{{\mathrm min}}=0\farcs035 \times 0\farcs023, PA=86^\circ$ using natural weight as {\it u-v} sampling,  which is almost the same as that of the 226 GHz continuum maps.  The sensitivities of the original velocity width channels, $\Delta V\simeq 20$ km s$^{-1}$, are $1\sigma=110~ \mu$Jy beam$^{-1}$ or  3.1 K in $T_\mathrm{B}$ in the emission-free areas. 
The observation parameters of the H30$\alpha$ recombination line  observation are also summarized in Table 1.

\subsection{Archival Data from Jun. 2019}
We have analyzed the archival data of Sgr A$^\ast$ itself and the surrounding area, which had been obtained in the ALMA Cy.6 season (2018.1.001124.S. PI L. Murchikova), in order to measure the proper motions of the objects detected in the Cy.5 observation. The major observation parameters,  namely the center position, diameter of the FOV,  and frequency resolution of the spectral window, are the same as those of the Cy.5 observation. The spectral window of $f_c=229.5$ GHz is for the continuum observation.
This observation was performed in five days within the duration from 12 Jun. 2019 to 21 Jun. 2019.
The significant flaring activity of Sgr A$^\ast$ had been observed in the duration (\cite{MurchikovaLena}). 
As mentioned in the previous subsection, such short-time scale variation of Sgr A$^\ast$ induces some image quality degradation in the resultant map. Accordingly, we used only the last two days data to minimize the effect.  The observation epoch is $t_\mathrm{obs}=2019.471$ year.  The integration times of Sgr A* and J1744-3116 in one scan were 36 s and 12 s, respectively. The duration of one scan was about 75 s. The total elapsed time of one observation was about 4500 s.
The imaging was done using CASA 6.0 with {\tt tclean} task. We used the ``self-calibration" method  to obtain a high dynamic range in the continuum map.  After 8 iterations which is the same as that of the Cy.5 observation, the resultant map is obtained.
The angular resolution using natural weight as {\it u-v} sampling is $\theta_{{\mathrm maj}}\times\theta_{{\mathrm min}}=0\farcs027 \times 0\farcs024, PA=72^\circ$. 
The sensitivity in the emission-free area is $1\sigma=39~ \mu$Jy beam$^{-1}$ or 1.4 K in $T_\mathrm{B}$.  Compared to the October 2017 data, the dynamic range is degraded to $\sim90000$ due to degradation of image quality caused by the flare activity in Sgr A$^\ast$.
The observation parameters of this observation are also summarized in Table 1.
\begin{table}
  \caption{Observation Parameters}
  \begin{center} 
  \label{ }
\begin{tabular}{lccc}
\hline
\hline 
Parameter&343 GHz observation&226 GHz observation&230 GHz  observation \\
\hline
Program&2015.1.01080.S.&2017.1.00503.S.& 2018.1.001124.S.$^5$\\
Epoch (year)&2016.693&2017.773&2019.471\\
Reduction software &CASA 5.5 &CASA 5.6&CASA 6.0\\
Center frequency of SPW [GHz]&337.5, 339.5, 347.5, 349.5 &217.5, 219.5, 231.9, 234.0&229.5\\
Frequency band width [GHz]& 2.0&2.0&2.0\\
Frequency channel width [MHz]&15.625&15.625&15.625\\
FOV [$\arcsec$]&17&26&26\\
Angular resolution, $PA$&  &  &\\
~~~NW$^1$&- &$0\farcs037 \times 0\farcs025, 85^{\circ}$&$0\farcs027 \times 0\farcs024, 72^{\circ}$\\
~~~NW(H30$\alpha)^1$&- &$0\farcs035 \times 0\farcs023, 86^{\circ~4}$&-\\
~~~BW$^2$&$0\farcs107 \times 0\farcs101, -80^\circ$  &$0\farcs029 \times 0\farcs020, 86^{\circ}$&- \\
~~~UW$^3$&$0\farcs102 \times 0\farcs096, 80^\circ$&-&-\\
Continuum sensitivity&  &  &\\
~~~NW$^1$&- &$12~ \mu$Jy beam$^{-1}$ &$39~ \mu$Jy beam$^{-1}$\\
~~~BW$^2$&$100~\mu$Jy beam$^{-1}$&$15~ \mu$Jy beam$^{-1}$&-\\
~~~UW$^3$&$80~\mu$Jy beam$^{-1}$&-&-\\
Line sensitivity (bin)&  &  &\\
~~~NW(H30$\alpha)^1$&-&$110~ \mu$Jy beam$^{-1}$ (20 km s$^{-1})$&-\\
\hline
\end{tabular}
\end{center}
$^1$ natural weight. 
$^2$ Briggs weight (robust parameter = 0.5). 
$^3$ uniform weight.
$^4$ typical value.
$^5$ We use the archival data.
\end{table}
\begin{figure}
\begin{center}
\includegraphics[width=17cm, bb=0 0 959 881]{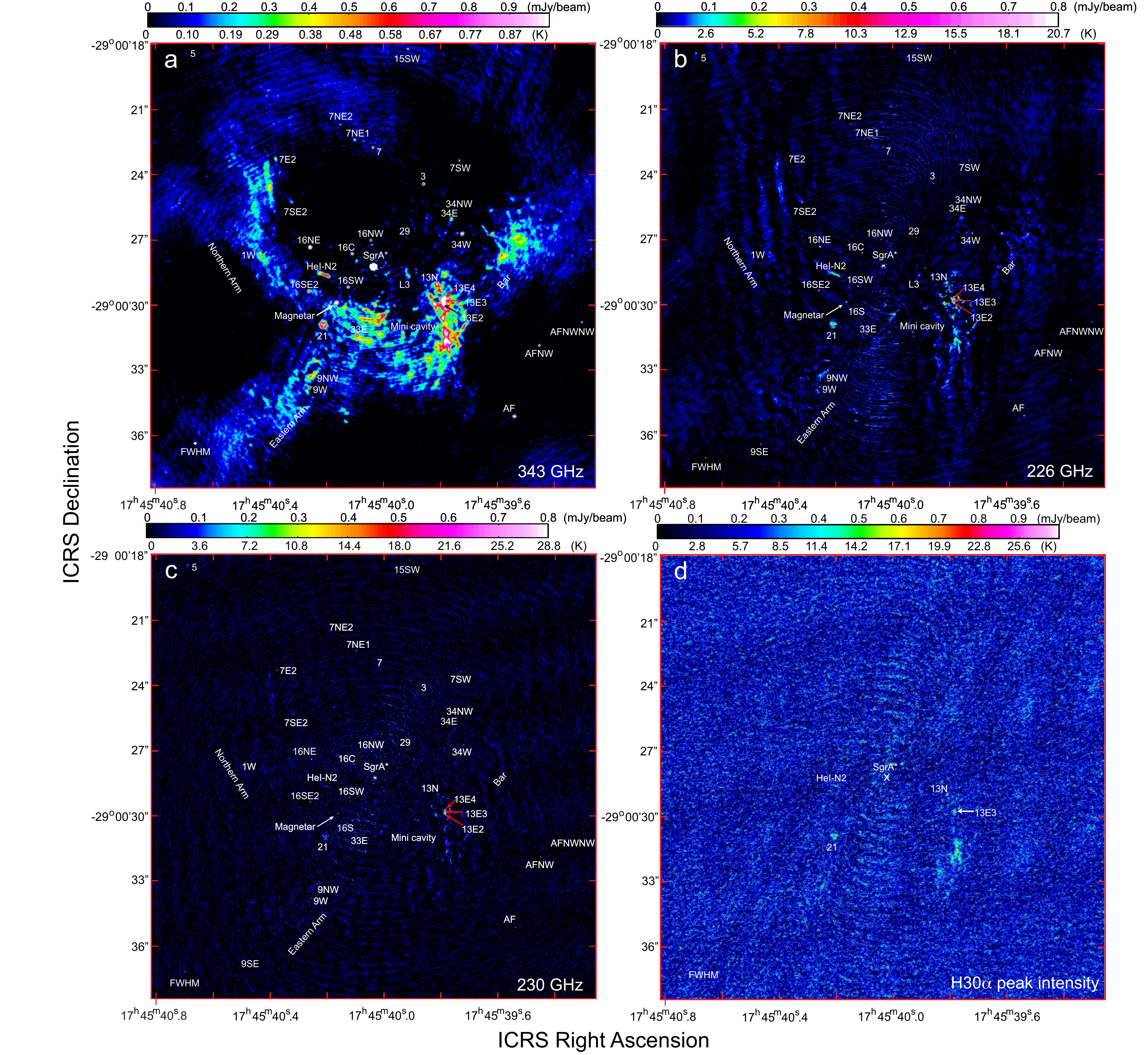}
\end{center}
 \caption{ {\bf a} Continuum map using Briggs weight at 343 GHz around Sgr A$^\ast$ in the ALMA Cy.3 program (2015.1.01080.S.)(see \cite{Tsuboi2017b}).  The angular resolution is $0\farcs107 \times 0\farcs101, PA=-78^\circ$  in FWHM, which is shown as an oval at the lower left corner.  This is a finding chart of the objects around Sgr A$^\ast$.
The position of of Sgr A$^\ast$ was $\alpha_{\rm ICRS}$ = $17^{\rm h}45^{\rm m}40^{\rm s}.033903 \pm0^{\rm s}.0000034$,$\delta_{\rm ICRS}$= $-29^{\circ}00'28\farcs224140\pm0\farcs000506$ at  $t_\mathrm{obs}=2016.693$ year using 2-D Gaussian fitting. The peak flux density of Sgr A$^\ast$ at 343 GHz is derived to be  $S_\nu{\mathrm{(343)}}=2182\pm27$ mJy beam$^{-1}$.
 {\bf b} Continuum map using natural weight at 226 GHz around Sgr A$^\ast$ in the ALMA Cy.5 program (2017.1.00503.S.) (also see \cite{Tsuboi2019}).  The angular resolution is $0\farcs037 \times 0\farcs025, PA=-86^\circ$  in FWHM, which is shown as an oval at the lower left corner. 
The position of Sgr A$^\ast$ was $\alpha_{\rm ICRS}$ = $17^{\rm h}45^{\rm m}40^{\rm s}.40.03338820\pm0^{\rm s}.00000025$, $\delta_{\rm ICRS}$= $-29^{\circ}00'28\farcs23047008\pm0\farcs00000147$ at  $t_\mathrm{obs}=2017.773$ year (see text). The peak flux density of Sgr A$^\ast$ at 226 GHz is derived to be  $S_\nu{\mathrm{(226)}}=3069\pm0.7$ mJy beam$^{-1}$.
{\bf c} Continuum map using natural weight at 230 GHz around Sgr A$^\ast$ in the ALMA Cy.6 program (2017.1.00503.S.)  The angular resolution is $\theta_\mathrm{maj} \times \theta_\mathrm{min}=0\farcs027 \times 0\farcs024, PA = 72^{\circ}$ in FWHM, which is shown as an oval at the lower left corner. 
The position of Sgr A$^\ast$ was $\alpha_{\rm ICRS}$ = $17^{\rm h}45^{\rm m}40^{\rm s}.0329918\pm0^{\rm s}.0000002$, $\delta_{\rm ICRS}$= $-29^{\circ}00'28\farcs242434\pm0\farcs000002$ at $t_\mathrm{obs}=2019.471$ year. The peak flux density of Sgr A$^\ast$ at 230 GHz is derived to be  $S_\nu{\mathrm{(226)}}=3569.8\pm0.7$ mJy beam$^{-1}$ at  $t_\mathrm{obs}=2019.471$.
 {\bf d} Peak intensity map  using natural weight around Sgr A$^\ast$ in the H$30\alpha$ recombination line in the ALMA Cy.5 program (2017.1.00503.S.). The angular resolution is $0\farcs035 \times 0\farcs023, PA=86^\circ$ in FWHM, which is shown as an oval at the lower left corner. The cross shows the position of Sgr A$^\ast$.
 }
 \label{Fig1}
\end{figure}
\begin{figure}
\begin{center}
\includegraphics[width=8cm, bb=0 0 517 726 ]{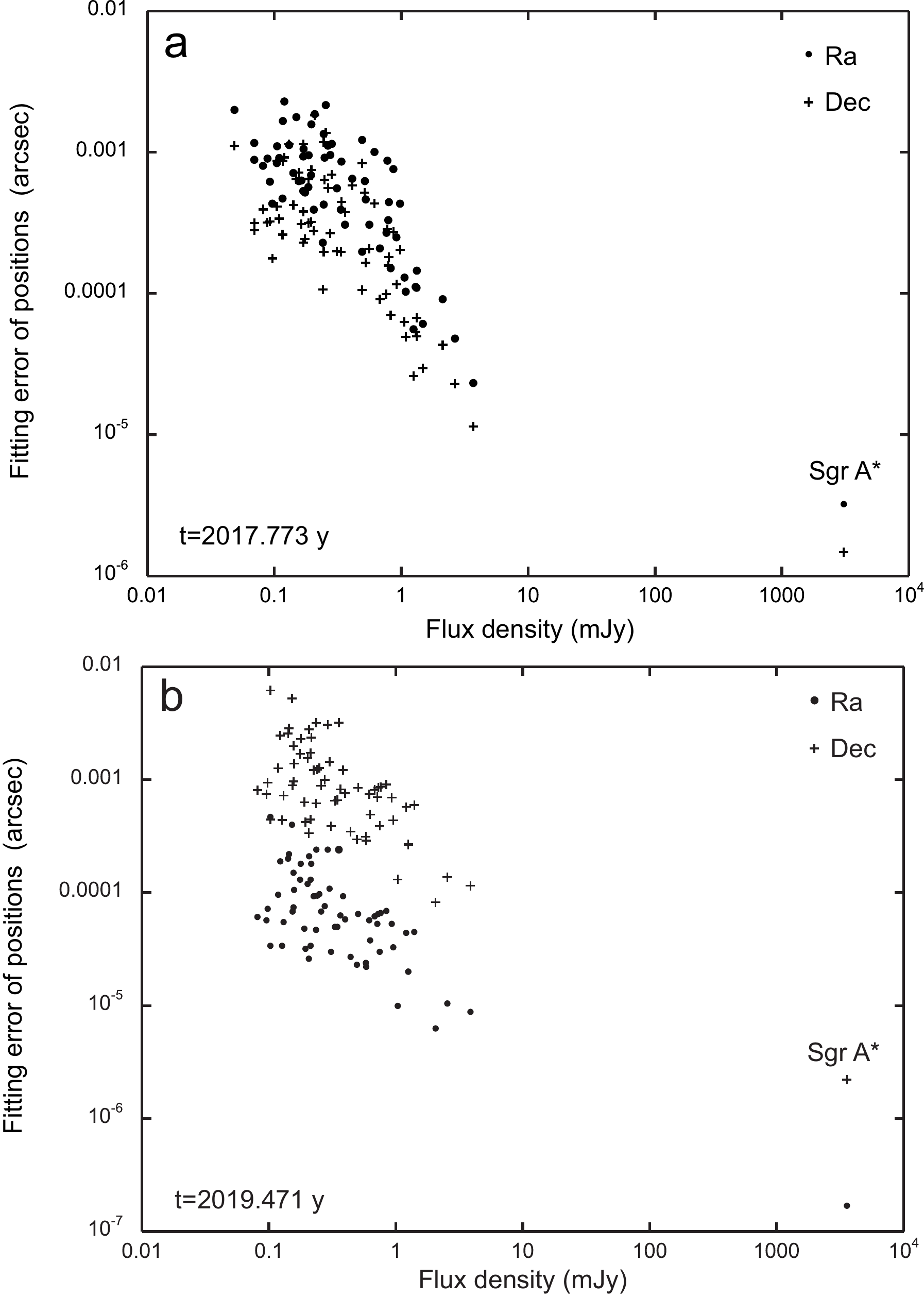}
\end{center}
 \caption{ Relation between the flux density of the detected source and the 2-D Gaussian fitting positional error. {\bf a} in the epoch of  2017.773 year. {\bf b}  in the epoch of  2019.471 year.}
 \label{Fig2}
\end{figure}
\begin{figure}
\begin{center}
\includegraphics[width=17cm, bb=0 0 917 418 ]{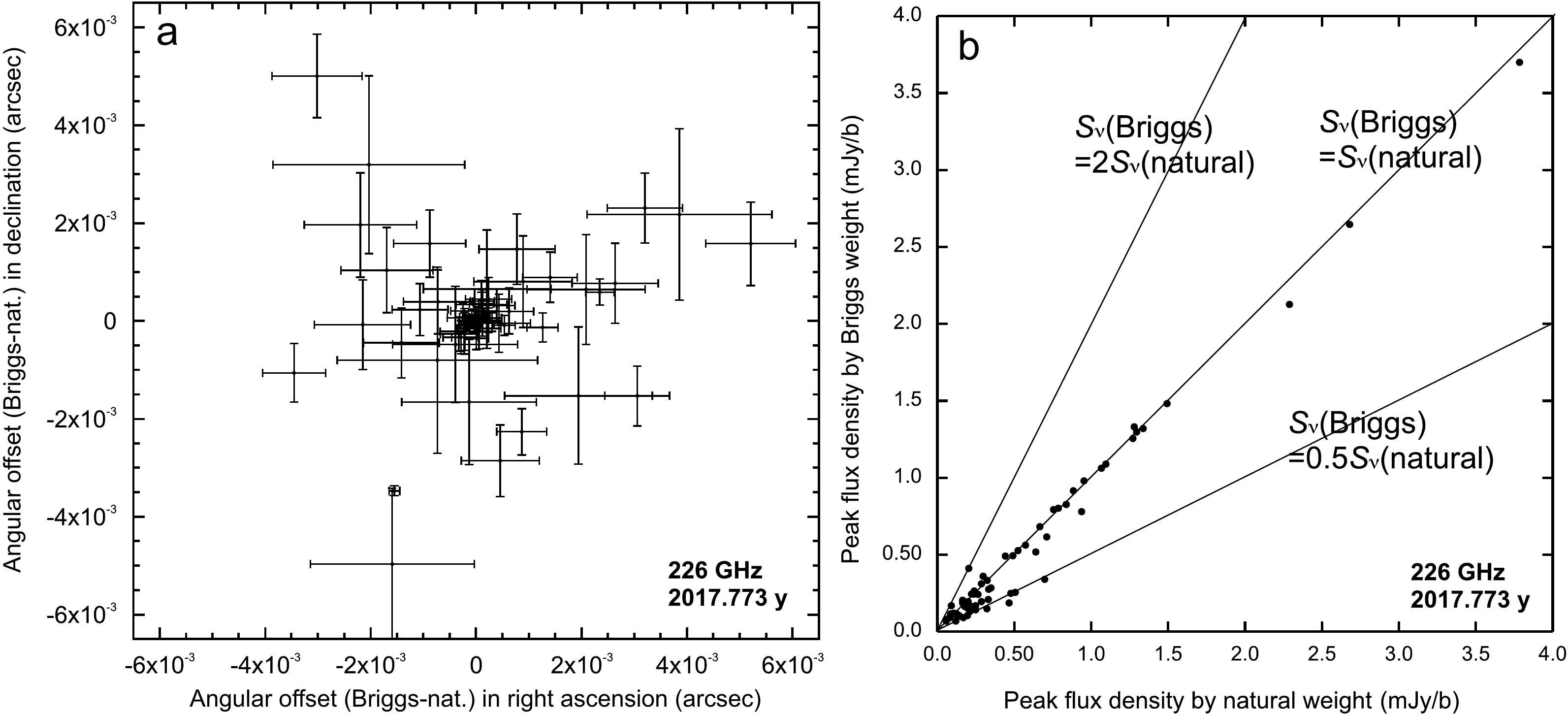}
\end{center}
 \caption{ {\bf a} Angular offsets between the star positions derived using Briggs and natural weights in the epochs of 2017.773 year. {\bf b} Relation between the flux densities derived using Briggs and natural weights in the epoch of  2017.773 year. }
 \label{Fig3}
\end{figure}

\section{Results} 
\subsection{Continuum Maps and Positions of the Detected Objects}
Figure 1a shows the continuum map  using Briggs weight of the  region within 0.4 pc of Sgr A$^\ast$ at 343 GHz obtained in the ALMA Cy.3 program (2015.1.01080.S.)(\cite{Tsuboi2017b}). The observation epoch is $t_{\mathrm{Cy.3}}=2016.693$ y. The angular resolution is $0\farcs107 \times 0\farcs101, PA=-78^\circ$  in FWHM.  Although Sgr A$^\ast$ is prominent in the figure, the GCMS which is composed of the Northern Arm(NA),  Eastern Arm(EA) and Bar are also identified clearly. Therefore this figure is a finding chart of the objects around Sgr A$^\ast$, which will be discussed. 

Figures 1b shows the continuum map using natural weight at 226 GHz around Sgr A$^\ast$ in the ALMA Cy.5 program (2017.1.00503.S.) (also see \cite{Tsuboi2019}).  The angular resolution is $0\farcs037 \times 0\farcs025, PA=-86^\circ$  in FWHM.  Sgr A$^\ast$ is dominant in the figure, while the GCMS is identified vaguely. 
The position of Sgr A$^\ast$ at 226 GHz was $\alpha_{\rm ICRS}$ = $17^{\rm h}45^{\rm m}40^{\rm s}.40.03338820\pm0^{\rm s}.00000025$, $\delta_{\rm ICRS}$= $-29^{\circ}00'28\farcs23047008\pm0\farcs00000147$ at $t_\mathrm{obs}=2017.773$ year using 2-D Gaussian fitting for the continuum map using Briggs weight (CASA task, {\tt imfit}\footnote{
The errors of central position, peak flux density, and angular size in this paper are estimated by the standard procedure of CASA (\cite{Condon}).
}).  Although the positions of Sgr A$^\ast$ using Briggs and natural weights are identical within the accuracy of $<1~\mu$as, we used the star positions with Briggs weight for the data of 2017.773 year in the following analysis because of the reason shown later. These positional errors include only fitting errors not calibration errors. In the case of Sgr A$^\ast$, the signal-to-noise ratio ($SNR$) is equal to the dynamic range mentioned in S2.1, $SNR\gtrsim2\times10^5$. The very high $SNR$ is thought to make the very high nominal accuracy (Cf. Figure 2a).
The peak flux density of Sgr A$^\ast$ at 226 GHz is derived to be  $S_\nu{\mathrm{(226)}}=3069\pm0.7$ mJy beam$^{-1}$ at $t_{\mathrm{obs}}=2017.773$ year using 2-D Gaussian fitting.  The absolute accuracy of the flux density should be as large as $10~\%$ based on the ALMA User Guide. 
The second prominent component in these figures is the IRS13E cluster which is centered around $\alpha_\mathrm{ICRS}\sim17^\mathrm{h}45^\mathrm{m}39.^\mathrm{s}8, \delta_\mathrm{ICRS}\sim-29^\circ00'29.\arcsec8$.  
The IRS13E cluster has a very complicated structure extending to the south. The southern extension has been identified as the ionized gas flowing in the IRS13E  cluster (\cite{Tsuboi2017b}). 
Other well-known IR objects around Sgr A$^\ast$ including, the IRS16 cluster and IRS21 and so on are also clearly detected both in the continuum map (also see Figure 1a). They are labeled in the figures. The positions and flux densities of these objects are derived by 2-D Gaussian fitting to the continuum map and summarized in Table 2. 
The angular extent of these objects is also estimated by 2-D Gaussian deconvolution. When the estimated extent is smaller than the beam size, the object is determined to be point-like. The total number of the detected objects around Sgr A$^\ast$ is 65. The detected objects are distributed in the angular distance from Sgr A$^\ast$ from $\sim1\arcsec$ to $\sim13\arcsec$.  The flux densities of the detected objects are distributed in the range from $\sim0.05$ mJy to $\sim4$ mJy.
Additionally, the positions, flux densities, and angular extent of these objects at 343 GHz are derived by the same procedure. They are also summarized in Table 2.

Figures 1c shows the continuum map using natural weight at at 230 GHz around Sgr A$^\ast$ in the ALMA Cy.6 program (2018.1.001124.S.).
The angular resolution is $0\farcs027 \times 0\farcs024, PA=72^\circ$  in FWHM. 
The position of Sgr A$^\ast$ was $\alpha_{\rm ICRS}$ = $17^{\rm h}45^{\rm m}40^{\rm s}.03299175\pm0^{\rm s}.00000017$, $\delta_{\rm ICRS}$= $-29^{\circ}00'28\farcs24243361\pm0\farcs00000180$ at $t_\mathrm{obs}=2019.471$ year using 2-D Gaussian fitting.
The peak flux density of Sgr A$^\ast$ at 230 GHz is derived to be  $S_\nu\mathrm{(226)}=3569.8\pm0.7$ mJy beam$^{-1}$ at  $t_\mathrm{obs}=2019.471$ using 2-D Gaussian fitting. 
The positions and flux densities of the objects around Sgr A$^\ast$ are also derived by 2D Gaussian fitting and summarized in Table 2. All the objects detected in the Cy.5 observation are detected except one object, No.64, which would be under the detection limit around the edge of the FOV.

As mentioned above, the positions of the detected objects are derived by 2-D Gaussian fitting to the continuum maps. We checked the relation between the flux density and the fitting positional error.  Figures 2a and 2b show the relations  observed at  2017.773 and 2019.471 years, respectively. The fitting positional error seems to decrease with increasing the flux density both in the epochs. This is thought to be caused by that  the fitting is affected by signal-to-noise ratio of the object.
In the epochs of 2017.773 year, the fitting positional  error is similar both in  right ascension and declination and up to a few mas (see Figure 2a).
 On the other hand, the error in the epochs of 2019.471 year is fairly different in the right ascension and declination (see Figure 2b). 
Although the error in right ascension is almost less than 1 mas, the error in declination often reaches to several mas. This situation may be caused especially by the artifact around Sgr A$^\ast$ mentioned above in the observation data at 2019.471 year. If the positional error of 1 mas is in these observations, this causes the ambiguity of $\sim30$ km s$^{-1}$ as proper motion. 

Figure 3a shows the angular offsets between the positions relative to Sgr A$^\ast$ derived using Briggs and natural weights in the epochs of 2017.773 year. The error bars in the figure indicate the fitting errors.
 Because the synthesized beam size (beam solid angle) using natural weight, is $60 \%$ larger than that using Briggs weight, the former is easily affected by the extended emission around the object.
Especially, if the extended emission is not uniform, the position derived by 2-D Gaussian fitting should be shifted from the original position.
Although the angular offsets of almost stars are less than $<2$ mas,  several stars have relatively large offset up to 5 mas. 
The stars with large offset have large fitting errors. Because these stars are, for the most part, located in the IRS13E and IRS13N clusters, the large offsets would be caused by complex extended emission around stars in the clusters (see Figure 9).  Therefore, it is better to use the star positions with Briggs weight for the data of 2017.773 year.  On the other hand, we used unavoidably the star positions with natural weight for the data of 2019.471 year because the stars with weak intensity are undetectable because of the insufficient $SNR$ if we use Briggs weight. However, because the beam with natural weight for the data of 2019.471 year, $0\farcs027 \times 0\farcs024$, is comparable with the beam with Briggs weight for the data of 2017.773 year, $0\farcs029 \times 0\farcs020$, this would not significantly affect the derivation of proper motions.
In addition,  Figure 3b shows the relation between the flux densities derived using Briggs and natural weights  in the epochs of 2017.773 year. As mentioned above,  if the beam solid angle is larger the measurement is more easily affected by the extended emission. There is a trend that the flux densities derived using natural weight are somewhat larger than those derived using Briggs weight. Especially, this trend is noticeable in the flux density less than 1 mJy. The trend may be caused by that larger beam solid angle samples larger extended emission.  This is another reason for that we used Briggs weight whenever possible. 

\begin{figure}
\begin{center}
\includegraphics[width=17cm, bb=0 0 830 819]{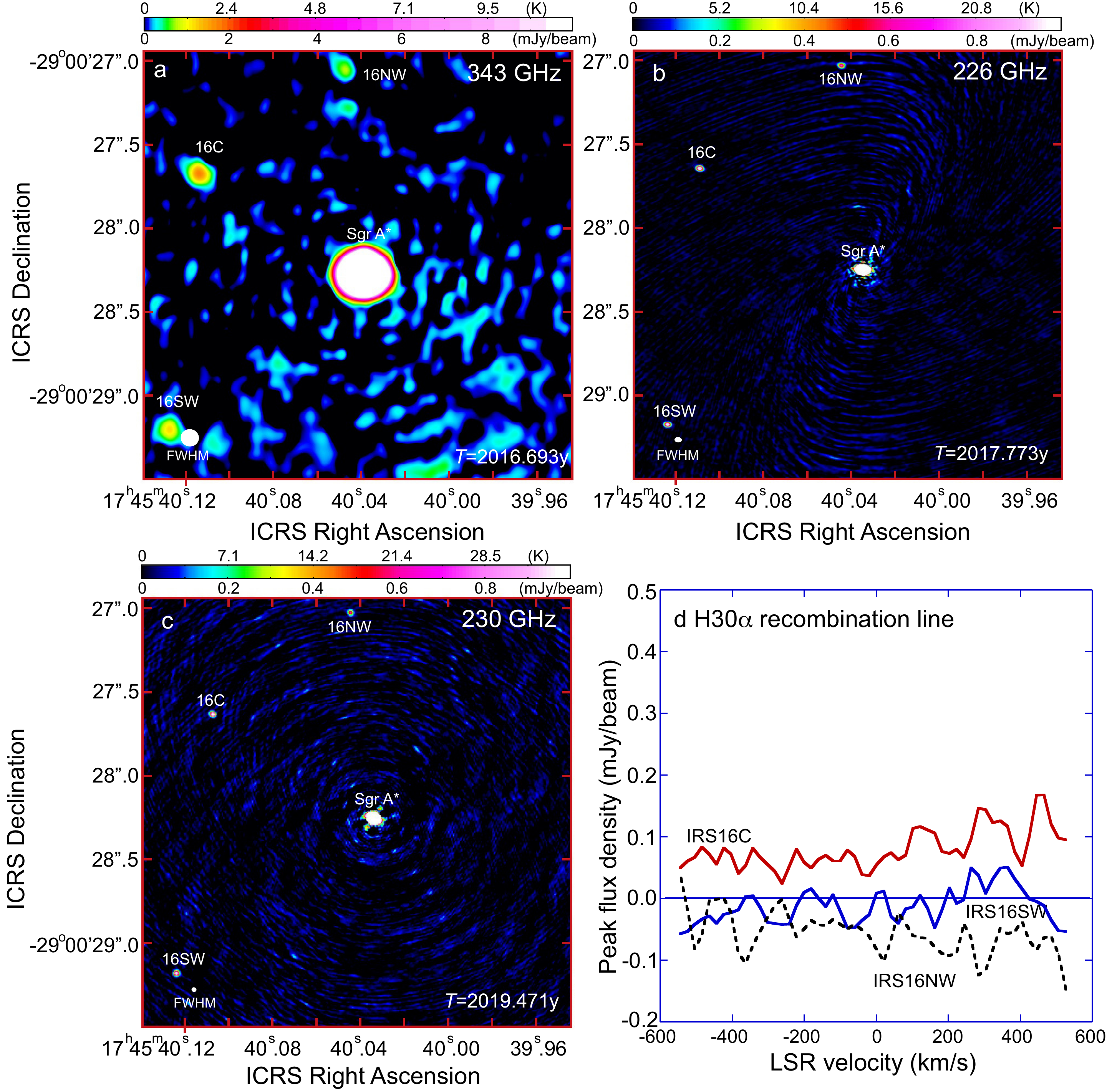}
\end{center}
 \caption{{\bf a} Enlarged continuum map using uniform weight in the vicinity within 0.1 pc of Sgr A$^\ast$ at 343 GHz. The angular resolution is $0\farcs102 \times 0\farcs096, PA=80^\circ$  in FWHM, which is shown as an oval at the lower left corner.   
The position of Sgr A$^\ast$ at 343 GHz  is $\alpha_{\rm ICRS}$ = $17^{\rm h}45^{\rm m}40^{\rm s}.033903\pm0^{\rm s}.000034$, $\delta_{\rm ICRS}$= $-29^{\circ}00'28\farcs224140\pm0\farcs000506$ at 2016.693 year. The peak flux density of Sgr A$^\ast$ is $2182\pm27$ mJy beam$^{-1}$ at 343 GHz.
 {\bf b} Enlarged continuum map  using Briggs weight in the same area at 226 GHz. The angular resolution is $0\farcs029 \times 0\farcs020, PA=86^\circ$  in FWHM, which is shown as an oval at the lower left corner. 
The position of Sgr A$^\ast$ at 226 GHz is $\alpha_{\rm ICRS}$ = $17^{\rm h}45^{\rm m}40^{\rm s}.03338820\pm0^{\rm s}.00000025$, $\delta_{\rm ICRS}$= $-29^{\circ}00'28\farcs23047001\pm0\farcs0000015$ at 2017.773 year. The peak flux density of Sgr A$^\ast$ is $S_\nu=3069\pm0.7$ mJy beam$^{-1}$ at 226 GHz (also see \cite{Tsuboi2017}, 2019).
 {\bf c} Enlarged continuum map  using natural weight in the same area at 230 GHz . The angular resolution is $0\farcs027 \times 0\farcs025, PA=72^\circ$  in FWHM, which is shown as an oval at the lower left corner. 
The position of Sgr A$^\ast$ at 230 GHz is $\alpha_{\rm ICRS}$ = $17^{\rm h}45^{\rm m}40^{\rm s}.03299175\pm0^{\rm s}.00000017$, $\delta_{\rm ICRS}$= $-29^{\circ}00'28\farcs2424336\pm0\farcs0.0000018$ at 2019.471 year. The peak flux density of Sgr A$^\ast$ is $S_\nu=3570\pm0.7$ mJy beam$^{-1}$ at 230 GHz (also see \cite{Tsuboi2017}, 2019).
 {\bf d} Line profiles of IRS16NW, 16C and 16SW in H30$\alpha$ recombination line, which were obtained in the ALMA Cy.5 program. }
\label{Fig4}
\end{figure}
\subsection{Region within 0.1 pc of Sgr A$^\ast$}
Figures 4a, 4b, and 4c show the enlarged continuum maps in the vicinity within 0.1 pc of Sgr A$^\ast$ at 343, 226 and 230 GHz, respectively. 
Three famous WR stars, IRS16NW, IRS16C, and IRS16SW, are detected clearly around Sgr A$^\ast$ although Sgr A$^\ast$ is dominant in these figures. 
In Figures 4b, and 4c, the extended component larger than the beam size is not detected around Sgr A$^\ast$ at these frequencies, or Sgr A$^\ast$ is detected as a point source. 
Accordingly, the physical extent of the emitting region of Sgr A$^\ast$ is thought to be much less than that of the beam size, $25~\mathrm{mas}\sim0\farcs001$ pc. 
Similarly, the extended components larger than the beam size are not detected around IRS16NW, IRS16C, and IRS16SW.

There are some faint patches, $S_\nu\sim0.2$ mJy beam$^{-1}$, in the immediate vicinity ($\lesssim0\farcs25\sim0.01$ pc) of Sgr A$^\ast$ in Figures 4b and 4c. These patches are probably residuals of the side lobes after applying the ``self-calibration" method.
If the patches are residuals of the side lobes, the dynamic range of the map is proved to reach to $\gtrsim100000$ immediately around Sgr A$^\ast$.
Although famous S stars are located in this area, the patches are probably not these stars because the distribution resembles to that of the side lobe not that of the stars. 
It is considered that the S stars are too faint to be detected in these observations because they have the spectral type of B. 

The peak flux densities of Sgr A$^\ast$ are $S_\nu=2.182\pm0.027$ Jy beam$^{-1}$ at 343 GHz  at 2016.693 year, $S_\nu=3.069\pm0.0007$ Jy beam$^{-1}$ at 226 GHz at 2017.773 year, and $S_\nu=3.570\pm0.0007$ Jy beam$^{-1}$ at 230 GHz  at 2019.471 year, respectively. The flux density of Sgr A$^\ast$ is variable temporally in the millimeter and sub-millimeter wavelengths as well known (e.g. \cite{Miyazaki}). Therefore, the spectral index cannot be derived using these data.

\subsection{H$30\alpha$ Recombination Line Map}
Figures 1d  shows the peak intensity map in the H$30\alpha$ recombination line obtained in the Cy.5 observation. 
The cross in the figure shows the position of Sgr A$^{\star}$. 
The widely extended structures of the GCMS are almost disappeared in Figures 1c. However,  IRS13E3 and the southern extension are still prominent. IRS21, HeI-N2, and IRS13N cluster are also identified although they are faint. 
The disappearance of the GCMS is considered to be caused by that the structure is mainly consist of extended components over the maximum detectable angular scale of the Cy.5 observation and they are  resolved-out.
Many IR stars around Sgr A$^\ast$ including IRS16NW, IRS16C, IRS16SW, and so on are thought to be WR stars, which should have ionized gas outflows.
The radial velocity derived from the H$30\alpha$ recombination line data is useful to obtain the 3-D kinematics.
However, such ionized gas outflows were not detected in the H$30\alpha$ recombination line data of the Cy.5 observation. 
Figure 4d shows the line profiles of IRS16NW, IRS16C, and IRS16SW in the H$30\alpha$ recombination line as examples of the non-detection.

\subsection{Flux densities of Member Stars of the Nuclear Star Cluster}
We checked the reproducibility in the observed flux densities.  Figure 5 shows the relation between the flux densities at 226 and 230 GHz,  which have been observed at  2017.773 and  2019.471 year, respectively. 
The detected objects are mostly WR stars. The centimeter-wave observations of WR stars  in the Galactic disk have been performed using VLA and ATCA (e.g. \cite{Cappa}, \cite{Andrews}). Although the intensity variabilities of some WR stars have been reported (e.g. \cite{Cappa}), it is an open question whether WR stars are usually variable in the timescale of less than several years or not.
We assumed here that WR stars are not variable significantly within the elapse time.
Although the flux densities at 2019.471 year are nearly equal to or slightly smaller than those at 2017.773 year in the intensity region of $S_\nu\gtrsim1.5$ mJy, the former are somewhat  smaller than the latter in the weaker intensity region. 
Moreover, the data points are scattered randomly in the intensity region of $S_\nu\lesssim0.4$ mJy.
The dependence on the intensity would indicate that the difference is caused by the reproducibility rather than time variation of the objects.
The degree of differences between observed flux densities would indicate the degree of reproducibility in the observed flux densities.
In addition, the random behavior in the intensity region of $S_\nu\lesssim0.4$ mJy may be also caused by the artifact mentioned previously in the observation data at 2019.471 year . 

There is a remaining issue what is the spectrum in mm-submm wavelengths of  detected objects.
This issue affects the capability of astrometry with ALMA. If many objects have spectra dominated by dusts or the spectral index is steeper  than $\alpha\gtrsim2$ ($S_\nu\propto \nu^\alpha$), shorter wavelength observations should detect many fainter objects ($S_\nu(343)\gtrsim 2 \times S_\nu(230)$). On the other hand, if many objects have spectra dominated by ionized gas or the spectral index is as flat as $\alpha\sim-0.1$, longer wavelength observations should be promising in order to detect many fainter objects ($FOV\sim60\arcsec$ at 90 GHz).
Although the spectrum indexes of WR stars in centimeter-wavelength are known to be widely scattered (\cite{Andrews}), those in mm-submm wavelengths are not yet studied comprehensively. 
The WR stars in the NSC may be good observation targets to solve the issue because they are located at the same distance from us. 
Figure 6 shows the comparison between flux densities of detected objects at 230 and 343 GHz. The flux densities at 230 GHz are the means of those at 226 and 230 GHz which are summarized in Table 1. The linear lines indicate the spectral indexes of $S_\nu\propto \nu^{-0.1}$, $S_\nu\propto \nu^1$ and $S_\nu\propto \nu^2$, respectively. 
Spectral indexes of the detected WR stars are scattered apparently in the  wide range from $\alpha\sim-0.1$ to $\alpha\sim2$.  However, it is impossible to determine if this is due to that they have such scattered intrinsic spectral indexes or the flux densities are variable in the observation duration ($\sim2.8$ years). 
Therefore it is inconclusive (from the data) whether shorter or longer wavelength observations are advantageous over one another.
\begin{figure}
\begin{center}
\includegraphics[width=9cm, bb=0 0 410 407]{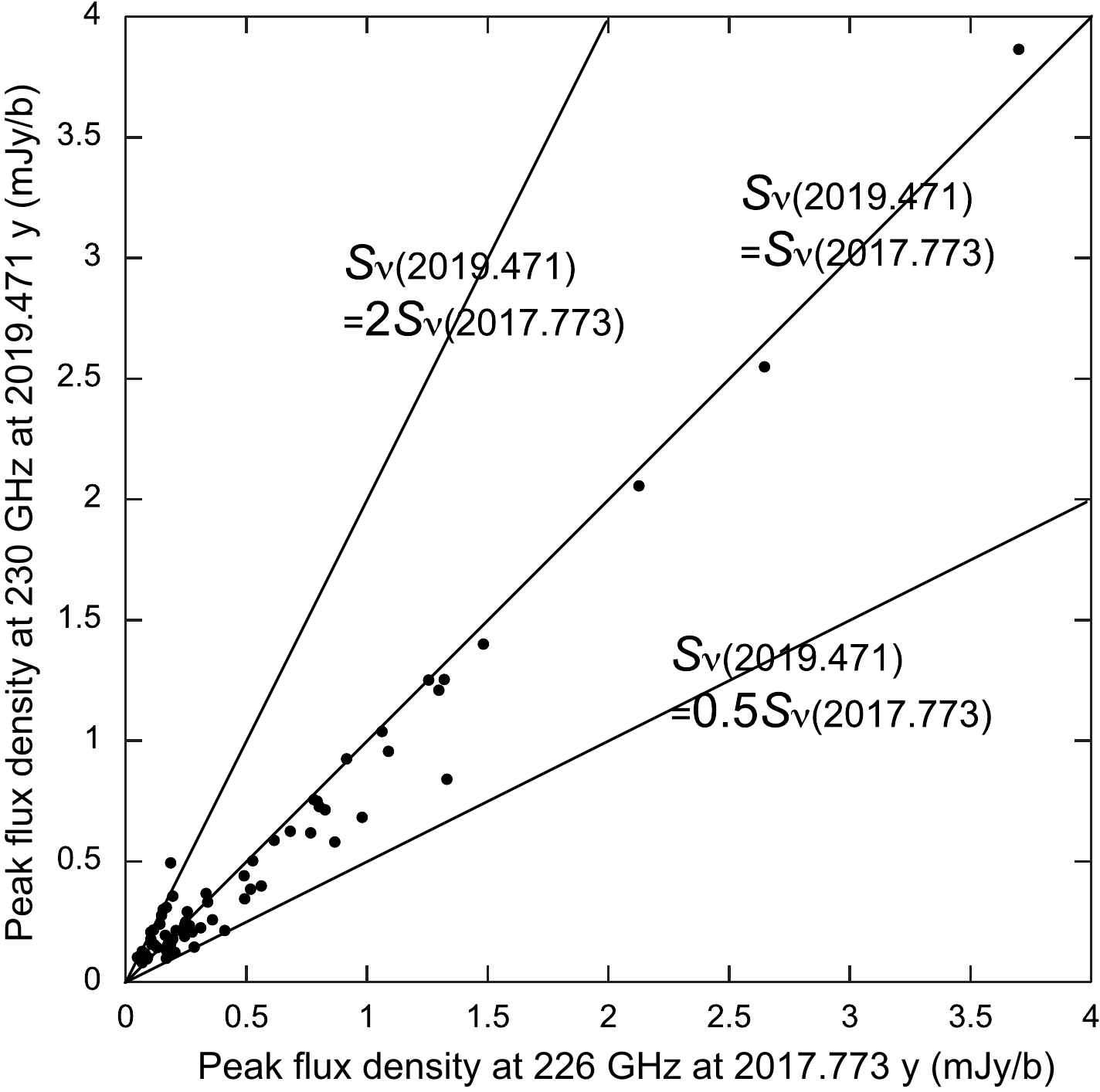}
\end{center}
 \caption{Relation between the flux densities at 226 and 230 GHz, which have been observed at  2017.773 and  2019.471 year, respectively.  Linear lines indicate the relations of $S_\nu(2019.471)=2S_\nu(2017.773)$, $S_\nu(2019.471)=S_\nu(2017.773)$ and $S_\nu(2019.471)=0.5S_\nu(2017.773)$, respectively. }
 \label{Fig5}
\end{figure}
\begin{figure}
\begin{center}
\includegraphics[width=9cm, bb=0 0 409 418 ]{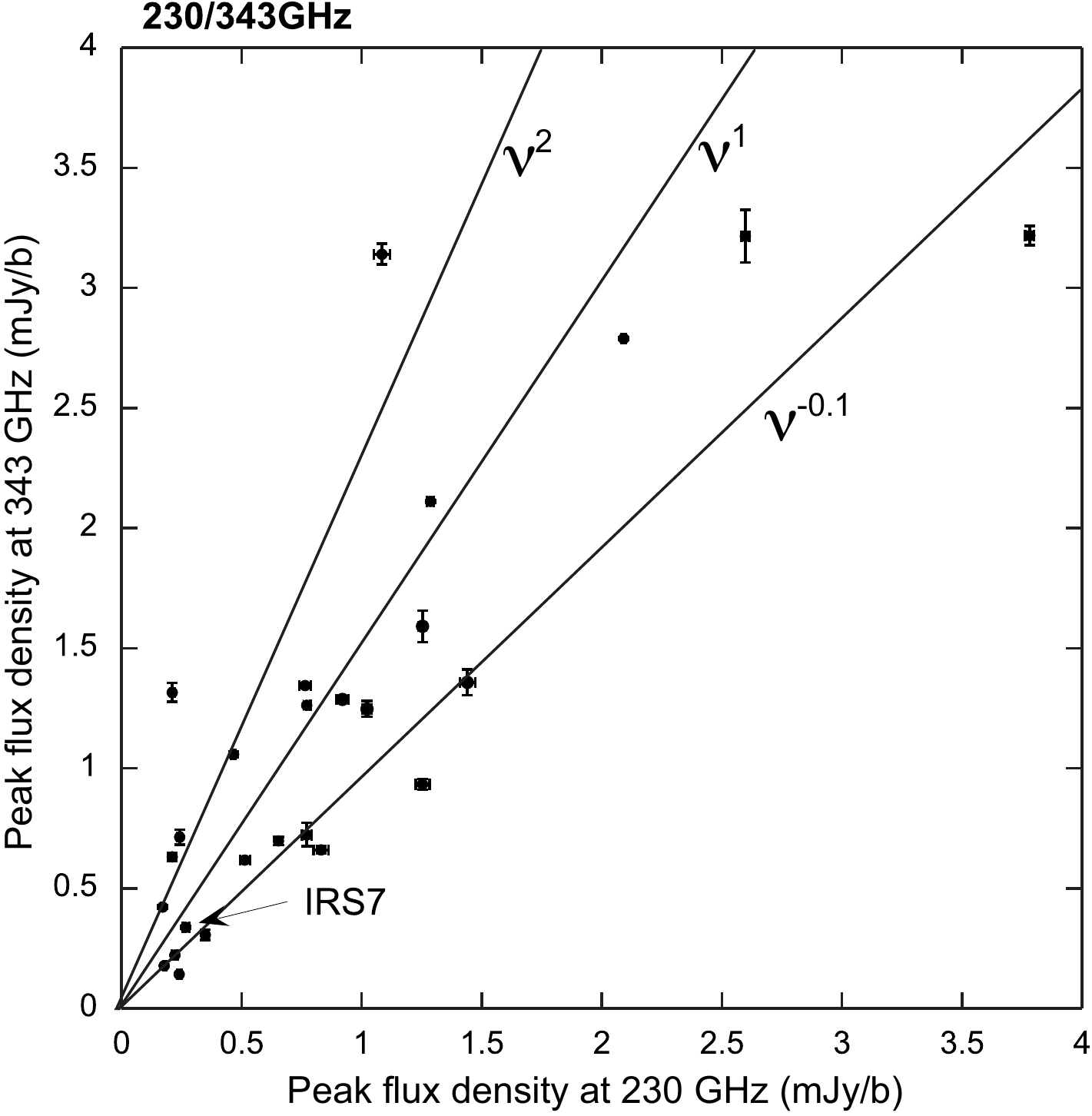}
\end{center}
 \caption{Comparison between flux densities of detected objects at 230 and 343 GHz. The flux densities at 230 GHz are the means of those at 226 and 230 GHz which are summarized in Table 2.  Linear lines indicate the spectra with $S_\nu\propto \nu^{-0.1}$, $S_\nu\propto \nu^1$ and $S_\nu\propto \nu^2$, respectively.}
 \label{Fig6}
\end{figure}

\clearpage
\begin{table}
  \caption{Positions, Flux Densities and Angular Extents of the Detected Objects around Sagittarius A$^\ast$ }\label{tab:first}
\begin{center}
\scalebox{0.72}[0.72] {
\begin{tabular}{lccccccc}
    \hline
    \hline
No.
&Name
&$\alpha_{\mathrm{Cy.3}} -17^{\rm h}45^{\rm m~1}$
&$\alpha_{\mathrm{Cy.5}} -17^{\rm h}45^{\rm m~2}$
&$\alpha_{\mathrm{Cy.6}} -17^{\rm h}45^{\rm m~3}$
&$-\delta_{\mathrm{Cy.3}}+29^{\circ}00'^{~1}$
&$-\delta_{\mathrm{Cy.5}}+29^{\circ}00'^{~2}$
&$-\delta_{\mathrm{Cy.6}}+29^{\circ}00'^{~3}$
\\
&
&[$^{\rm s}$]
&[$^{\rm s}$]
&[$^{\rm s}$]
&[$\arcsec$]
&[$\arcsec$]
&[$\arcsec$]
\\
\hline
1&Sgr A$^\ast$&40.033903$\pm$0.000034&40.03338820$\pm$0.00000025&40.03299175$\pm$0.00000017
&28.224140$\pm$0.000506&28.23047008$\pm$0.0000015&28.24243361$\pm$0.0000018
\\
2&IRS16C&40.10919$\pm$0.00016&40.10779$\pm$0.00001&40.10623$\pm$0.00005
&27.62910$\pm$0.00205&27.62384$\pm$0.00003&27.62363$\pm$0.00043
\\
3&IRS16NW&40.04378$\pm$0.00028&40.04312$\pm$0.00001&40.04354$\pm$0.00005
&27.00140$\pm$0.00434&27.00745$\pm$0.00007&27.01943$\pm$0.00056
\\
4&IRS16SW&40.12218$\pm$0.00008&40.12243$\pm$0.00001&40.12275$\pm$0.00004
&29.15942$\pm$0.00113&29.15940$\pm$0.00005&29.16482$\pm$0.00047
\\
5&IRS29&39.91689$\pm$0.00030&39.91802$\pm$0.00003&39.91802$\pm$0.00007
&26.86776$\pm$0.00362&26.90544$\pm$0.00104&26.92769$\pm$0.00062
\\
6&IRS16CC&-&40.18365$\pm$0.00005&40.18278$\pm$0.00018
&-&27.58607$\pm$0.00032&27.59042$\pm$0.00137
\\
7&IRS16SE1&-&40.18123$\pm$0.00004&40.18208$\pm$0.00018
&-&29.31932$\pm$0.00026&29.31145$\pm$0.00255
\\
8&IRS16S&-&40.14119$\pm$0.00004&40.14220$\pm$0.00007
&-&30.11295$\pm$0.00024&30.12517$\pm$0.0009
\\
9&Magnetar&40.16418$\pm$0.00006&40.16359$\pm$0.00001&40.16335$\pm$0.00007
&29.89075$\pm$0.00069&29.88759$\pm$0.00007&29.88852$\pm$0.00088
\\
10&IRS13N$\alpha$U&-&39.83333$\pm$0.00009&39.83429$\pm$0.00020
&-&29.54574$\pm$0.00115&29.54607$\pm$0.00303
\\
11&IRS13N$\alpha$L&-&39.83231$\pm$0.00009&39.83123$\pm$0.00006
&-&29.65141$\pm$0.00028&29.65500$\pm$0.00039
\\
12&IRS13N$\delta$&-&39.81219$\pm$0.00013&39.81167$\pm$0.00040
&-&28.99155$\pm$0.00087&28.99629$\pm$0.00304
\\
13&IRS13N$\epsilon$&-&39.81175$\pm$0.00005&39.81186$\pm$0.00047
&-&29.15795$\pm$0.00032&29.15393$\pm$0.00697
\\
14&IRS16NE&40.25621$\pm$0.00005&40.25585$\pm$0.00001&40.25583$\pm$0.00001
&27.33097$\pm$0.00058&27.34910$\pm$0.00001&27.37654$\pm$0.00010
\\
15&IRS33SW&-&40.01189$\pm$0.00007&40.01039$\pm$0.00015
&-&31.30012$\pm$0.00034&31.30508$\pm$0.00047
\\
16&IRS16SE2&40.26062$\pm$0.00010&40.26065$\pm$0.00001&40.26061$\pm$0.00003
&29.35163$\pm$0.00134&29.34682$\pm$0.00005&29.35070$\pm$0.00032
\\
17&IRS33E&40.09114$\pm$0.00008&40.09116$\pm$0.00003&40.09169$\pm$0.00006
&31.37651$\pm$0.00145&31.37890$\pm$0.00021&31.39145$\pm$0.00109
\\
18&IRS13Ef&-&39.82013$\pm$0.00002&39.81910$\pm$0.00007
&-&29.90060$\pm$0.00038&29.90373$\pm$0.00203
\\
19&IRS13N$\gamma$&-&39.79274$\pm$0.00008&39.79232$\pm$0.00021
&-&29.12088$\pm$0.00087&29.12138$\pm$0.00310
\\
20&IRS13E6&-&39.79663$\pm$0.00017&39.79456$\pm$0.00007
&-&29.70879$\pm$0.00092&29.72235$\pm$0.00189
\\
21&IRS13E1&-&39.80485$\pm$0.00006&39.80268$\pm$0.00009
&-&29.93819$\pm$0.00042&29.92984$\pm$0.00082
\\
22&IRS13Ea&-&39.78680$\pm$0.00009&39.78515$\pm$0.00022
&-&29.65390$\pm$0.00070&29.67294$\pm$0.00235
\\
23&IRS13Eb&-&39.78750$\pm$0.00005&39.78676$\pm$0.00009
&-&29.71359$\pm$0.00052&29.72144$\pm$0.00154
\\
24&IRS13E3&39.78884$\pm$0.00003&39.78929$\pm$0.00007&39.78863$\pm$0.00007
&29.76710$\pm$0.00047&29.75799$\pm$0.00029&29.76952$\pm$0.00049
\\
25&IRS13E4&39.78455$\pm$0.00002&39.78294$\pm$0.00003&39.78163$\pm$0.00007
&29.64059$\pm$0.00024&29.63148$\pm$0.00018&29.64326$\pm$0.00061
\\
26&IRS13Ec&-&39.78738$\pm$0.00002&39.78656$\pm$0.00006
&-&29.78667$\pm$0.00010&29.80080$\pm$0.00051
\\
27&IRS21-LR&-&40.20498$\pm$0.00016&40.20405$\pm$0.00024
&-&31.03001$\pm$0.00137&31.04889$\pm$0.00230
\\
28&-&-&39.78932$\pm$0.00007&39.78926$\pm$0.00005
&-&29.87323$\pm$0.00045&29.88445$\pm$0.00045
\\
29&IRS13Ed&-&39.78608$\pm$0.00015&39.78545$\pm$0.00005
&-&29.85531$\pm$0.00167&29.87340$\pm$0.00117
\\
30&IRS13E2&39.78666$\pm$0.00001&39.78542$\pm$0.00001&39.78415$\pm$0.00001&
29.93966$\pm$0.00009&29.95779$\pm$0.00009&29.96853$\pm$0.00007
\\
31&IRS13E5&-&39.77233$\pm$0.00005&39.77133$\pm$0.00011
&-&29.67291$\pm$0.00072&29.68135$\pm$0.00246
\\
32&-&-&39.76755$\pm$0.00005&39.76611$\pm$0.00003
&-&29.75879$\pm$0.00058&29.76175$\pm$0.00005
\\
33&IRS13Ee&-&39.78124$\pm$0.00007&39.78060$\pm$0.00010
&-&30.14124$\pm$0.00064&30.14738$\pm$0.00144
\\
34&-&-&39.75869$\pm$0.00004&39.75748$\pm$0.00007
&-&29.92265$\pm$0.00023&29.92835$\pm$0.00049
\\
\hline
\end{tabular}}\end{center}
$^1$ The observation epoch is $t_{\mathrm{Cy.3}}=2016.693$ y, 
$^2$ The observation epoch is $t_{\mathrm{Cy.5}}=2017.773$ y. 
$^3$ The observation epoch is $t_{\mathrm{Cy.6}}=2019.471$ y.
$^4$ When the deconvolved image size is much smaller than the synthesized beam ($30$ mas at 230 GHz, $40$ mas at 226 GHz and $100$ mas at 343 GHz), we define that it is a point source, P. 
$^5$ The peak flux density is derived by 2D Gaussian fit.
\end{table}
\clearpage
\addtocounter{table}{-1}
\begin{table}
  \caption{Continued.}\label{tab:first}
\begin{center}
\scalebox{0.72}[0.72] {
\begin{tabular}{lccccccc}
    \hline
    \hline
No.
&Name
&$\alpha_{\mathrm{Cy.3}} -17^{\rm h}45^{\rm m~1}$
&$\alpha_{\mathrm{Cy.5}} -17^{\rm h}45^{\rm m~2}$
&$\alpha_{\mathrm{Cy.6}} -17^{\rm h}45^{\rm m~3}$
&$-\delta_{\mathrm{Cy.3}}+29^{\circ}00'^{~1}$
&$-\delta_{\mathrm{Cy.5}}+29^{\circ}00'^{~2}$
&$-\delta_{\mathrm{Cy.6}}+29^{\circ}00'^{~3}$
\\
&
&[$^{\rm s}$]
&[$^{\rm s}$]
&[$^{\rm s}$]
&[$\arcsec$]
&[$\arcsec$]
&[$\arcsec$]
\\
\hline
35&-&-&39.78882$\pm$0.00007&39.78793$\pm$0.00002&-&30.63723$\pm$0.00065&30.64490$\pm$0.00044\\
36&-&-&39.76084$\pm$0.00015&39.75933$\pm$0.00003&-&25.99978$\pm$0.00111&25.99888$\pm$0.00049\\
37&IRS34W&39.72258$\pm$0.00001&39.72198$\pm$0.00001&39.72130$\pm$0.00002&26.71637$\pm$0.00013&26.72453$\pm$0.00005&26.74586$\pm$0.00021\\
38&IRS3&39.85890$\pm$0.00022&39.85866$\pm$0.00007&39.85885$\pm$0.00003
&24.42073$\pm$0.00296&24.41246$\pm$0.00027&24.42189$\pm$0.00023\\
39&IRS7SE2&-&40.26501$\pm$0.00006&40.26533$\pm$0.00013
&-&24.76677$\pm$0.00041&24.77751$\pm$0.00187\\
40&IRS13W&39.69455$\pm0.00001$&39.69530$\pm$0.00005&39.69425$\pm$0.00003
&29.82434$\pm0.00025$&29.85065$\pm$0.00031&29.86040$\pm$0.00035\\
41&IRS34NW&39.742516$\pm$0.00006&39.74234$\pm$0.00004&39.74091$\pm$0.00011
&25.45558$\pm0.00138$&25.43943$\pm$0.00032&25.45611$\pm$0.00090\\
42&-&39.718828$\pm$0.00001&39.72013$\pm$0.00014&39.71996$\pm$0.00008&
31.04740$\pm0.00025$&31.05306$\pm$0.00065&31.05975$\pm$0.00087\\
43&-&-&39.76859$\pm$0.00008&39.76793$\pm$0.00003
&-&32.05287$\pm$0.00115&32.04728$\pm$0.00058\\
44&IRS6E&-&39.64166$\pm$0.00003&39.64096$\pm$0.00019
&-&27.37381$\pm$0.00028&27.37978$\pm$0.00342\\
45&IRS7&40.03641$\pm$0.00003&40.03569$\pm$0.00004&40.03572$\pm$0.00009&
22.73652$\pm$0.00098&22.76113$\pm$0.00020&22.78018$\pm$0.00132\\
46&IRS7NE1&40.09975$\pm$0.00005&40.09938$\pm$0.00002&40.09862$\pm$0.00003&
22.38461$\pm$0.00058&22.38118$\pm$0.00011&22.38746$\pm$0.00030\\
47&IRS7SW&39.73254$\pm$0.00004&39.73235$\pm$0.00004&39.73176$\pm$0.00007&
23.33843$\pm$0.00080&23.35198$\pm$0.00017&23.36886$\pm$0.00096\\
48&IRS9NW&40.25797$\pm0.00001$&40.25736$\pm$0.00002&40.25786$\pm$0.00005&
33.78795$\pm0.00007$&33.79589$\pm$0.00012&33.80294$\pm$0.00053\\
49&IRS7W&39.85270$\pm$0.00005&39.85306$\pm$0.00010&39.85308$\pm$0.00024&
22.24258$\pm$0.00050&22.23176$\pm$0.00118&22.24105$\pm$0.00239\\
50&IRS9W&40.26824$\pm$0.00024&40.26429$\pm$0.00014&40.26421$\pm$0.00013&
33.95782$\pm$0.00219&33.93974$\pm$0.00183&33.94384$\pm$0.00144\\
51&IRS7E2&40.37711$\pm$0.00008&40.37628$\pm$0.00002&40.37660$\pm$0.00004&
23.27175$\pm$0.00110&23.27927$\pm$0.00009&23.29230$\pm$0.00037\\
52&IRS7NE2&40.15053$\pm$0.00002&40.14897$\pm$0.00004&40.14792$\pm$0.00012&
21.68862$\pm$0.00033&21.69063$\pm$0.00021&21.69902$\pm$0.00108\\
53&IRS6W&-&39.44574$\pm$0.00007&39.44569$\pm$0.00006
&-&26.76967$\pm$0.00038&26.76884$\pm$0.00043\\
54&IRS6NW&-&39.445294$\pm$0.00007&39.44469$\pm$0.00003
&-&26.51045$\pm$0.00032&26.52010$\pm$0.00041\\
55&AFNW&39.45235$\pm$0.00005&39.45158$\pm$0.00001&39.45100$\pm$0.00002&
31.84201$\pm$0.00064&31.85837$\pm$0.00003&31.87609$\pm$0.00023\\
56&-&-&40.03150$\pm$0.00006&40.03133$\pm$0.00010
&-&19.64663$\pm$0.00040&19.65594$\pm$0.00163\\
57&IRS10E&-&40.61884$\pm$0.00007&40.61920$\pm$0.00006
&-&24.05270$\pm$0.00032&24.07079$\pm$0.00062\\
58&-&-&40.75190$\pm$0.00003&40.75113$\pm$0.00005
&-&27.98623$\pm$0.00020&28.00458$\pm$0.00069\\
59&AF&39.53955$\pm$0.00002&39.53929$\pm$0.00001&39.53925$\pm$0.00001
&35.11251$\pm$0.00032&35.11620$\pm$0.00002&35.12552$\pm$0.00011\\
60&AFNWNW&39.30302$\pm$0.00001&39.30261$\pm$0.00003&39.30239$\pm$0.00003
&30.79032$\pm$0.00012&30.79656$\pm$0.00016&30.81048$\pm$0.00033\\
61&IRS9SE&-&40.46340$\pm$0.00002&40.46280$\pm$0.00006
&-&36.43745$\pm$0.00021&36.45220$\pm$0.00056\\
62&-&-&39.37495$\pm$0.00009&39.37430$\pm$0.00005
&-&33.49673$\pm$0.00056&33.50216$\pm$0.00044\\
63&IRS15SW&-&39.91264$\pm$0.00008&39.91200$\pm$0.00002
&-&18.21589$\pm$0.00044&18.23050$\pm$0.00021\\
64&-&-&39.35564$\pm$0.00003&-
&-&33.95517$\pm$0.00018&-\\
65&-&-&40.13729$\pm$0.00001&40.13680$\pm$0.00001
&-&16.52472$\pm$0.00006&16.53267$\pm$0.00011\\
66&IRS5&-&40.69051$\pm$0.00012&40.69057$\pm$0.00024
&-&18.43462$\pm$0.00075&18.45965$\pm$0.00375\\
\hline
\end{tabular}}
\end{center}
$^1$ The observation epoch is $t_{\mathrm{Cy.3}}=2016.693$ y, 
$^2$ The observation epoch is $t_{\mathrm{Cy.5}}=2017.773$ y. 
$^3$ The observation epoch is $t_{\mathrm{Cy.6}}=2019.471$ y.
$^4$ When the deconvolved image size is much smaller than the synthesized beam ($30$ mas at 230 GHz, $40$ mas at 226 GHz and $100$ mas at 343 GHz), we define that it is a point source, P. 
$^5$ The peak flux density is derived by 2D Gaussian fit.
\end{table}
\clearpage
\addtocounter{table}{-1}
\begin{table}
  \caption{Continued.}\label{tab:first}
\begin{center}
\scalebox{0.72}[0.72] {
\begin{tabular}{lcccccc}
    \hline
    \hline
No.
&Size$_{\mathrm{Cy.3}}$, $PA^{~1,4}$
&Size$_{\mathrm{Cy.5}}$, $PA^{~2,4}$
&Size$_{\mathrm{Cy.6}}$, $PA^{~3,4}$
&$S_\nu(343)^{1,5}$
&$S_\nu(226)^{2,5}$
&$S_\nu(230)^{3,5}$
\\
&[mas$\times$mas, deg.]
&[mas$\times$mas, deg.]
&[mas$\times$mas, deg.]
& [mJy beam$^{-1}$]
& [mJy beam$^{-1}$]
& [mJy beam$^{-1}$]
\\
\hline
1&P&P&P&2182$\pm$27&3069$\pm$0.7&3570$\pm$0.7\\
2&P&P&P&1.358$\pm$0.053&1.483$\pm$0.006&1.400$\pm$0.065\\
3&P&P&P&0.724$\pm$0.049&0.828$\pm$0.008&0.710$\pm$0.040\\
4&P&P&P&0.938$\pm$0.022&1.298$\pm$0.010&1.209$\pm$0.056\\
5&P&P&P&0.307$\pm$0.022&0.335$\pm$0.009&0.366$\pm$0.022\\
6&-&P&P&-&0.196$\pm$0.009&0.178$\pm$0.028\\
7&-&P&P&-&0.182$\pm$0.004&0.215$\pm$0.022\\
8&-&P&P&-&0.175$\pm$0.005&0.157$\pm$0.016\\
9&P&P&P&3.141$\pm$0.044&1.331$\pm$0.013&0.840$\pm$0.064\\
10&-&$56\times27, ~161$&$50\times10, ~142$&-&0.132$\pm$0.011&0.142$\pm$0.019\\
11&-&P&P&-&0.070$\pm$0.004&0.081$\pm$0.005\\
12&-&$116\times23, ~121$&$58\times26, ~112$&-&0.117$\pm$0.007&0.153$\pm$0.028\\
13&-&P&$57\times25, ~39$&-&0.092$\pm$0.004&0.103$\pm$0.027\\
14&P&P&P&3.219$\pm$0.040&3.700$\pm$0.006&3.865$\pm$0.037\\
15&-&$50\times22, 95$&$70\times14, 93$&-&0.109$\pm$0.004&0.156$\pm$0.008\\
16&P&P&P&1.248$\pm$0.032&1.089$\pm$0.007&0.955$\pm$0.034\\
17&P&P&P&0.660$\pm$0.015&0.981$\pm$0.025&0.683$\pm$0.056\\
18&-&$48\times33, ~172$&$48\times17, ~6$&-&0.361$\pm$0.006&0.258$\pm$0.020\\
19&-&P&$46\times14, ~141$&-&0.106$\pm$0.006&0.207$\pm$0.031\\
20&-&$49\times19, ~106$&P&-&0.121$\pm$0.004&0.153$\pm$0.015\\
21&-&P&P&-&0.142$\pm$0.007&0.241$\pm$0.011\\
22&-&P&P&-&0.284$\pm$0.019&0.144$\pm$0.022\\
23&-&$49\times34, ~131$&$36\times26, ~5$&-&0.519$\pm$0.014&0.384$\pm$0.032\\
24&P&P&$41\times19, ~100$&5.053$\pm$0.036&0.780$\pm$0.029&0.756$\pm$0.046\\
25&$96\times33, ~133$&P&P&1.345$\pm$0.007&0.803$\pm$0.022&0.727$\pm$0.046\\
26&-&$68\times29, ~90$&$38\times26, ~80$&-&0.766$\pm$0.006&0.618$\pm$0.022\\
27&-&48$\times$26, 119&44$\times$31, 111&-&0.255$\pm$0.022&0.291$\pm$0.040\\
28&-&52$\times$25, 112&P&-&0.340$\pm$0.011&0.331$\pm$0.014\\
29&-&$41\times32, ~106$&$32\times11, ~173$&-&0.494$\pm$0.027&0.346$\pm$0.021\\
30&P&P&P&2.790$\pm$0.005&2.127$\pm$0.013&2.057$\pm$0.014\\
31&-&P&$37\times15, ~11$&-&0.156$\pm$0.007&0.301$\pm$0.036\\
32&-&$49\times41, ~44$&P&-&0.412$\pm$0.012&0.213$\pm$0.009\\
33&-&P&P&-&0.249$\pm$0.013&0.249$\pm$0.023\\
34&-&P&P&-&0.170$\pm$0.004&0.098$\pm$0.006\\
\hline
\end{tabular}}
\end{center}
$^1$ The observation epoch is $t_{\mathrm{Cy.3}}=2016.693$ y, 
$^2$ The observation epoch is $t_{\mathrm{Cy.5}}=2017.773$ y. 
$^3$ The observation epoch is $t_{\mathrm{Cy.6}}=2019.471$ y.
$^4$ When the deconvolved image size is much smaller than the synthesized beam ($30$ mas at 230 GHz, $40$ mas at 226 GHz and $100$ mas at 343 GHz), we define that it is a point source, P. 
$^5$ The peak flux density is derived by 2D Gaussian fit.
\end{table}
\clearpage
\addtocounter{table}{-1}
\begin{table}
  \caption{Continued.}\label{tab:first}
\begin{center}
\scalebox{0.72}[0.72] {
\begin{tabular}{lcccccc}
    \hline
    \hline
No.
&Size$_{\mathrm{Cy.3}}$, $PA^{~1,4}$
&Size$_{\mathrm{Cy.5}}$, $PA^{~2,4}$
&Size$_{\mathrm{Cy.6}}$, $PA^{~3,4}$
&$S_\nu(343)^{1,5}$
&$S_\nu(226)^{2,5}$
&$S_\nu(230)^{3,5}$
\\
&[mas$\times$mas, deg.]
&[mas$\times$mas, deg.]
&[mas$\times$mas, deg.]
& [mJy beam$^{-1}$]
& [mJy beam$^{-1}$]
& [mJy beam$^{-1}$]
\\
\hline
35&-&P&P&-&0.187$\pm$0.007&0.494$\pm$0.012\\
36&-&P&P&-&0.049$\pm$0.004&0.103$\pm$0.005\\
37&P&P&P&2.112$\pm$0.005&1.320$\pm$0.009&1.253$\pm$0.027\\
38&$126\times92, ~168$&P&P&0.714$\pm$0.031&0.277$\pm$0.013&0.207$\pm$0.007\\
39&-&P&P&-&0.105$\pm$0.007&0.177$\pm$0.022\\
40&P&P&P&0.178$\pm0.001$&0.164$\pm$0.008&0.194$\pm$0.007\\
41&P&P&P&$0.423\pm0.007$&0.186$\pm$0.008&0.158$\pm$0.011\\
42&$161\times117$, 168&P&$35\times19$, 131&1.316$\pm0.039$&0.149$\pm$0.010&0.276$\pm$0.015\\
43&-&P&$64\times36$, 158&-&0.171$\pm$0.011&0.309$\pm$0.006\\
44&-&P&$41\times10$, 30&-&0.205$\pm$0.005&0.123$\pm$0.020\\
45&P&P&P&0.339$\pm$0.004&0.312$\pm$0.010&0.225$\pm$0.025\\
46&$173\times137, ~137$&P&P&1.058$\pm$0.014&0.493$\pm$0.007&0.440$\pm$0.014\\
47&P&P&P&0.619$\pm$0.010&0.526$\pm$0.014&0.503$\pm$0.040\\
48&P&P&P&1.287$\pm0.002$&0.915$\pm$0.015&0.926$\pm$0.048\\
49&P&P&P&0.143$\pm$0.003&0.245$\pm$0.019&0.236$\pm$0.023\\
50&$236\times153, ~69$&P&$31\times17, ~52$&0.632$\pm$0.018&0.209$\pm$0.027&0.214$\pm$0.021\\
51&P&P&P&0.697$\pm$0.017&0.682$\pm$0.010&0.625$\pm$0.025\\
52&P&P&P&0.222$\pm$0.002&0.235$\pm$0.006&0.202$\pm$0.025\\
53&-&P&P&-&0.169$\pm$0.009&0.131$\pm$0.006\\
54&-&P&P&-&0.070$\pm$0.005&0.127$\pm$0.004\\
55&P&P&P&1.591$\pm$0.065&1.256$\pm$0.005&1.252$\pm$0.029\\
56&-&P&P&-&0.082$\pm$0.005&0.118$\rm$0.012\\
57&-&P&P&-&0.089$\pm$0.004&0.096$\pm$0.005\\
58&-&P&P&-&0.246$\pm$0.006&0.190$\pm$0.011\\
59&P&P&P&3.217$\pm$0.109&2.647$\pm$0.009&2.549$\pm$0.030\\
60&P&P&P&1.263$\pm$0.003&0.794$\pm$0.018&0.749$\pm$0.025\\
61&-&P&P&-&0.564$\pm$0.014&0.398$\pm$0.026\\
62&-&P&P&-&0.265$\pm$0.018&0.234$\pm$0.012\\
63&-&P&P&-&0.616$\pm$0.038&0.586$\pm$0.013\\
64&-&P&-&-&0.096$\pm$0.003&-\\
65&-&P&P&-&1.063$\pm$0.009&1.039$\pm$0.012\\
66&-&P&P&-&0.196$\pm$0.011&0.356$\pm$0.017\\
\hline
\end{tabular}}
\end{center}
$^1$ The observation epoch is $t_{\mathrm{Cy.3}}=2016.693$ y, 
$^2$ The observation epoch is $t_{\mathrm{Cy.5}}=2017.773$ y. 
$^3$ The observation epoch is $t_{\mathrm{Cy.6}}=2019.471$ y.
$^4$ When the deconvolved image size is much smaller than the synthesized beam ($30$ mas at 230 GHz, $40$ mas at 226 GHz and $100$ mas at 343 GHz), we define that it is a point source, P. 
$^5$ The peak flux density is derived by 2D Gaussian fit.
\end{table}

\clearpage
\begin{table}
  \caption{Proper Motion of the Detected Objects around Sagittarius A$^\ast$}\label{tab:second}
   \begin{center}
   \scalebox{0.8}[0.8] {
\begin{tabular}{lccccccccc}
\hline\hline
No.&Name&$\Delta Ra^1$&$\Delta Dec^1$&$V_\mathrm{Ra}^{~~~2}$&$V_\mathrm{Ra,IR}$&$V_\mathrm{Dec}^{~~~3}$&$V_\mathrm{Dec,IR}$&$V_\bot^{~~4}$&$PA^{~5}$\\
&&(mas)&(mas)&(km s$^{-1}$)&(km s$^{-1}$)&(km s$^{-1}$)&(km s$^{-1}$)&(km s$^{-1}$)&(deg.)\\
\hline
1&Sgr A$^\ast$&0&0&0&-&0&-&0&-\\
2&IRS16C&-15.32$\pm$0.60&12.17$\pm$0.43&-342$\pm$13&-342$\pm50^6$&272$\pm$10&302$\pm44^6$&437$\pm$16&51.5$\pm$1.5\\
3&IRS16NW&10.78$\pm$0.72&0.08$\pm$0.56&241$\pm$16&$195\pm5^7$&2$\pm$13&$40\pm2^7$&241$\pm$20&-89.6$\pm$3.0\\
4&IRS16SW&9.33$\pm$0.58&6.54$\pm$0.48&208$\pm$13&$246\pm4^7$&146$\pm$11&$110\pm3^7$&254$\pm$17&-55.0$\pm$2.6\\
5&IRS29&5.16$\pm$0.91&-10.29$\pm$0.65&115$\pm$20&$148\pm4^7$&-230$\pm$14&$-202\pm5^7$&257$\pm$25&-153.4$\pm$4.3\\
6&IRS16CC&-6.25$\pm$2.40&7.62$\pm$1.41&-140$\pm$54&-99$\pm3^7$&170$\pm$31&249$\pm1^7$&220$\pm$62&39.4$\pm$12.0\\
7&IRS16SE1&16.42$\pm$2.42&19.83$\pm$2.56&366$\pm$54&184$\pm43^6$&443$\pm$57&124$\pm44^6$&575$\pm$79&-39.6$\pm$5.5\\
8&IRS16S&18.41$\pm$1.10&-0.26$\pm$1.00&411$\pm$24&301$\pm47^6$&-6$\pm$22&1$\pm43^6$&411$\pm$33&-90.8$\pm$3.1\\
9&Magnetar&2.03$\pm$0.92&11.04$\pm$0.88&45$\pm$20&-&246$\pm$20&-&251$\pm$28&-10.4$\pm$4.7\\
10&IRS13N$\alpha$U&17.83$\pm$2.81&11.63$\pm$3.24&398$\pm$63&(44$\pm8^8)^9$&260$\pm$72&(147$\pm13^8)^9$&475$\pm$96&-56.9$\pm$8.4\\
11&IRS13N$\alpha$L&-8.93$\pm$1.42&8.37$\pm$0.48&-199$\pm$32&(44$\pm8^8)^9$&187$\pm$11&(147$\pm13^8)^9$&273$\pm$33&46.9$\pm$4.8\\
12&IRS13N$\delta$&-1.62$\pm$5.51&7.22$\pm$3.16&-36$\pm$123&-33$\pm13^8$&161$\pm$71&248$\pm9^8$&165$\pm$142&12.6$\pm$42.0\\
13&IRS13N$\epsilon$&6.59$\pm$6.21&15.98$\pm$6.98&147$\pm$139&-58$\pm15^8$&357$\pm$156&306$\pm13^8$&386$\pm$209&-22.4$\pm$21.0\\
14&IRS16NE&5.00$\pm$0.12&-15.48$\pm$0.10&112$\pm$3&104$\pm49^6$&-346$\pm$2&-379$\pm47^6$&363$\pm$3&-162.1$\pm$0.4\\
15&IRS33SW&-14.52$\pm$2.19&7.00$\pm$0.58&-324$\pm$49&-&156$\pm$13&-&360$\pm$51&64.2$\pm3.9$\\
16&IRS16SE2&4.72$\pm$0.45&8.08$\pm$0.32&105$\pm$10&72$\pm5^7$&180$\pm$7&180$\pm2^7$&209$\pm$12&-30.3$\pm$2.6\\
17&IRS33E&12.23$\pm$0.92&-0.59$\pm$1.11&273$\pm$21&182$\pm47^6$&-13$\pm$25&-9$\pm42^6$&273$\pm$32&-92.8$\pm$5.2\\
18&IRS13Ef&3.43$\pm$0.94&8.83$\pm$2.06&77$\pm$21&-&197$\pm$46&-&211$\pm$51&-21.2$\pm$7.0\\
19&IRS13N$\gamma$&-0.27$\pm$3.01&11.46$\pm$3.22&-6$\pm$67&-114$\pm14^8$&256$\pm$72&298$\pm7^8$&256$\pm$98&1.3$\pm$15.0\\
20&IRS13E6&-22.01$\pm$2.46&-1.59$\pm$2.10&-491$\pm$55&-81$\pm25^7$&-36$\pm$47&97$\pm17^7$&492$\pm$72&94.1$\pm$5.5\\
21&IRS13E1&-23.32$\pm$1.42&20.31$\pm$0.92&-520$\pm$32&-195$\pm8^7$&453$\pm$21&-102$\pm5^7$&690$\pm$38&48.9$\pm$2.2\\
22&IRS13Ea&-16.50$\pm$3.07&-7.07$\pm$2.45&-368$\pm$69&-&-158$\pm$55&-&401$\pm$88&113.2$\pm$8.2\\
23&IRS13Eb&-4.56$\pm$1.37&4.11$\pm$1.62&-102$\pm$31&-&92$\pm$36&-&137$\pm$47&48.0$\pm$14.1\\
24&IRS13E3&-3.55$\pm$1.23&0.43$\pm$0.57&-79$\pm$27&-105$\pm23^7$&10$\pm$13&-16$\pm27^7$&80$\pm$30&83.0$\pm$9.3\\
25&IRS13E4&-11.85$\pm$0.96&0.19$\pm$0.64&-265$\pm$22&-254$\pm7^7$&4$\pm$14&15$\pm11^7$&265$\pm$26&89.1$\pm$3.1\\
26&IRS13Ec&-5.49$\pm$0.79&-2.16$\pm$0.51&-123$\pm$18&-&-48$\pm$11&-&132$\pm$21&111.5$\pm$5.4\\
27&IRS21-LR&-7.00$\pm$3.76&-6.92$\pm$2.68&-156$\pm$84&-&-154$\pm$60&-&220$\pm$103&134.7$\pm$19.0\\
28&-&4.36$\pm$1.08&0.743$\pm$0.63&97$\pm$24&-&17$\pm$14&-&99$\pm$28&-80.3$\pm$8.4\\
29&IRS13Ed&-2.95$\pm$1.39&-6.36$\pm$1.44&-66$\pm$31&-&-142$\pm$32&-&157$\pm$45&155.2$\pm$11.5\\
30&IRS13E2&-11.46$\pm$0.12&1.22$\pm$0.08&-256$\pm$3&-276$\pm6^7$&27$\pm$2&56$\pm8^7$&257$\pm$3&83.9$\pm$0.4\\
31&IRS13E5&-7.89$\pm$1.57&3.52$\pm$2.57&-176$\pm$35&-175$\pm45^8$&79$\pm$57&140$\pm72^8$&193$\pm$67&66.0$\pm$16.1\\
32&-&-13.68$\pm$0.79&8.99$\pm$0.80&-305$\pm$18&-&201$\pm$18&-&365$\pm$25&56.7$\pm$2.8\\
33&IRS13Ee&-3.22$\pm$1.57&5.82$\pm$1.57&-72$\pm$35&-&130$\pm$35&-&149$\pm$50&28.9$\pm$13.5\\
34&-&-10.61$\pm$1.08&6.27$\pm$0.54&-237$\pm$24&-&140$\pm$12&-&275$\pm$27&59.4$\pm$3.3\\
\hline 
\end{tabular}}
\end{center}
$^1$  $\Delta Ra$ and $\Delta Dec$ are the angular shifts between the relative positions at 2017.773 and 2019.471 year, which are the positions of Sgr A$^\ast$ subtracted from the star positions. $^2$ $V_\mathrm{Ra}=d\frac{\Delta Ra}{\Delta t}$, $d=8.0$ kpc. $^3$ $V_\mathrm{Dec}=d\frac{\Delta Dec}{\Delta t}$. $^4$ $V_\bot=\sqrt{V_\mathrm{Dec}^2+V_\mathrm{Ra}^2}$. $^5$ $PA=-\arctan{\frac{V_\mathrm{Ra}}{V_\mathrm{Dec}}}$.
$^6$ cited from \cite{Paumard}. $^7$ cited from \cite{Schodel2009}. $^8$ cited from \cite{Muzic2008}. $^9$ not resolved in IR.
\end{table}
\addtocounter{table}{-1}
\clearpage
\begin{table}
  \caption{Continued.}\label{tab:second}
   \begin{center}
   \scalebox{0.8}[0.8] {
\begin{tabular}{lccccccccc}
\hline\hline
No.&Name&$\Delta Ra^1$&$\Delta Dec^1$&$V_\mathrm{Ra}^{~~~2}$&$V_\mathrm{Ra,IR}$&$V_\mathrm{Dec}^{~~~3}$&$V_\mathrm{Dec,IR}$&$V_\bot^{~~4}$&$PA^{~5}$\\
&&(mas)&(mas)&(km s$^{-1}$)&(km s$^{-1}$)&(km s$^{-1}$)&(km s$^{-1}$)&(km s$^{-1}$)&(deg.)\\
\hline
35&-&-6.53$\pm$1.00&4.29$\pm$0.78&-146$\pm$226&-&96$\pm$17&-&174$\pm$28&56.7$\pm$6.2\\
36&-&-14.59$\pm$2.05&12.87$\pm$1.21&-326$\pm$46&-&287$\pm$27&-&434$\pm$53&48.6$\pm$4.8\\
37&IRS34W&-3.65$\pm$0.29&-9.36$\pm$0.22&-81$\pm$6&-79$\pm28^6$&-209$\pm$5&-116$\pm27^6$&224$\pm8$&158.7$\pm$1.6\\
38&IRS3&7.60$\pm$1.02&2.53$\pm$0.35&170$\pm$23&180$\pm6^7$&56$\pm$8&31$\pm4^7$&179$\pm$24&-71.6$\pm$3.3\\
39&IRS7SE2&9.42$\pm$1.90&1.23$\pm$1.92&210$\pm$42&250$\pm28^6$&27$\pm$43&32$\pm26^6$&212$\pm$60&-82.6$\pm$11.5\\
40&IRS13W&-8.60$\pm$0.76&2.21$\pm$0.47&-192$\pm$17&-180$\pm6^7$&49$\pm$10&64$\pm7^7$&198$\pm$20&75.6$\pm$3.2\\
41&IRS34NW&-13.61$\pm$1.50&-4.71$\pm$0.95&-304$\pm$33&-225$\pm28^6$&-105$\pm$21&-112$\pm27^6$&322$\pm$40&109.1$\pm$4.1\\
42&-&3.02$\pm$2.04&5.27$\pm$1.08&67$\pm$46&-&118$\pm$24&-&136$\pm$52&-29.8$\pm$17.4\\
43&-&-3.38$\pm$1.13&17.56$\pm$1.28&-75$\pm$25&-&392$\pm$29&-&399$\pm$38&10.9$\pm$3.6\\
44&IRS6E&-4.01$\pm$2.50&6.00$\pm$3.43&-89$\pm$56&-&134$\pm$77&-&161$\pm$95&33.8$\pm$22.4\\
45&IRS7&2.76$\pm$1.34&-7.08$\pm$1.33&62$\pm$30&-2$\pm9^7$&-158$\pm$30&-173$\pm12^7$&170$\pm$42&-158.7$\pm$10.1\\
46&IRS7NE1&-4.73$\pm$0.40&5.69$\pm$0.31&-106$\pm$9&-104$\pm4^7$&127$\pm$7&127$\pm5^7$&165$\pm$11&39.8$\pm$2.9\\
47&IRS7SW&-2.63$\pm$0.97&-4.92$\pm$0.98&-59$\pm$22&-5$\pm27^6$&-110$\pm$22&-108$\pm26^6$&124$\pm$31&151.8$\pm$10.0\\
48&IRS9NW&11.81$\pm$0.74&4.91$\pm$0.54&264$\pm$17&-&110$\pm$12&-&286$\pm$206&-67.5$\pm$2.6\\
49&IRS7W&5.53$\pm$3.46&2.68$\pm$2.66&123$\pm$77&185$\pm29^6$&60$\pm$59&36$\pm28^6$&137$\pm$98&-64.2$\pm$26.4\\
50&IRS9W&4.15$\pm$2.56&7.86$\pm$2.33&93$\pm$57&167$\pm29^6$&176$\pm$52&135$\pm27^6$&198$\pm$77&-27.8$\pm$16.1\\
51&IRS7E2&9.36$\pm$0.53&-1.07$\pm$0.38&209$\pm$12&177$\pm6^7$&-24$\pm$9&-36$\pm5^7$&210$\pm$15&-96.5$\pm$2.4\\
52&IRS7NE2&-9.32$\pm$1.58&3.31$\pm$1.09&-208$\pm$35&-178$\pm5^7$&74$\pm$24&114$\pm5^7$&221$\pm$43&70.4$\pm$6.7\\
53&IRS6W&4.54$\pm$1.18&12.79$\pm$0.57&101$\pm$26&-&286$\pm$13&-&303$\pm$29&-19.6$\pm$4.8\\
54&IRS6NW&-2.72$\pm$0.99&2.31$\pm$0.52&-61$\pm$22&-&52$\pm$12&-&80$\pm$25&49.6$\pm$12.1\\
55&AFNW&-2.38$\pm$0.27&-5.75$\pm$0.23&-53$\pm$6&-58$\pm6^7$&-128$\pm$5&-130$\pm4^7$&139$\pm$8&157.5$\pm$2.5\\
56&-&2.94$\pm$1.50&2.65$\pm$1.67&66$\pm$33&-&59$\pm$37&-&88$\pm$50&-48.0$\pm$23.1\\
57&IRS10E&10.03$\pm$1.17&-6.13$\pm$0.70&224$\pm$26&-&-137$\pm$16&-&262$\pm$30&-121.4$\pm$4.2\\
58&-&-4.81$\pm$0.77&-6.39$\pm$0.71&-107$\pm$17&-&-143$\pm$16&-&179$\pm$23&143.0$\pm$5.4\\
59&AF&4.69$\pm$0.15&2.64$\pm$0.12&105$\pm$3&80$\pm5^7$&59$\pm$3&44$\pm4^7$&120$\pm$4&-60.6$\pm$1.3\\
60&AFNWNW&2.38$\pm$0.51&-1.96$\pm$0.36&53$\pm$11&54$\pm4^7$&-44$\pm$8&-44$\pm4^7$&69$\pm$14&-129.5$\pm$8.0\\
61&IRS9SE&-2.59$\pm$0.82&-2.79$\pm$0.59&-58$\pm$18&-32$\pm4^7$&-62$\pm$13&-58$\pm4^7$&85$\pm$23&137.1$\pm$10.9\\
62&-&-3.26$\pm$1.28&6.54$\pm$0.71&-73$\pm$29&-96$\pm11^7$&146$\pm$16&99$\pm9^7$&163$\pm$33&26.5$\pm$9.3\\
63&IRS15SW&-3.23$\pm$1.05&-2.65$\pm$0.49&-72$\pm$23&-&-59$\pm$11&-&93$\pm$26&129.3$\pm$10.5\\
64&-&-&-&-&-&-&-&-&-\\				
65&-&-1.18$\pm$0.18&4.02$\pm$0.13&-26$\pm$4&3$\pm7^7$&90$\pm$3&73$\pm10^7$&93$\pm$5&16.4$\pm$2.5
\\
66&IRS5&5.99$\pm$3.57&-13.07$\pm$3.82&134$\pm$80&-&-292$\pm$85&-&321$\pm$117&-155.4$\pm$14.4\\
\hline
\end{tabular}}
\end{center}
$^1$  $\Delta Ra$ and $\Delta Dec$ are the angular shifts between the relative positions at 2017.773 and 2019.471 year, which are the positions of Sgr A$^\ast$ subtracted from the star positions. $^2$ $V_\mathrm{Ra}=d\frac{\Delta Ra}{\Delta t}$, $d=8.0$ kpc. $^3$ $V_\mathrm{Dec}=d\frac{\Delta Dec}{\Delta t}$. $^4$ $V_\bot=\sqrt{V_\mathrm{Dec}^2+V_\mathrm{Ra}^2}$. $^5$ $PA=-\arctan{\frac{V_\mathrm{Ra}}{V_\mathrm{Dec}}}$.
$^6$ cited from \cite{Paumard}. $^7$ cited from \cite{Schodel2009}. $^8$ cited from \cite{Muzic2008}.  $^9$ not resolved in IR.
\end{table}

\section{Proper Motions of Member Stars of the Nuclear Star Cluster}
\subsection{Proper Motion Referenced to Sgr A$^\ast$ of the Detected Objects in the Nuclear Star Cluster}
We derive the proper motions referenced to Sgr A$^\ast$ of the detected objects in the NSC based on the star position data at 2017.773 and 2019.471 year shown in Table 2 using the following procedure. 
First, the position at 2017.773 year of Sgr A$^\ast$, which is mentioned in the previous subsection,  is subtracted from the star positions of the epoch in order to calculate the relative positions referenced to Sgr A$^\ast$.
The star positions at 2019.471 year are also processed by the same procedure to calculate another relative positions.
The angular shifts between these derived relative positions, $\Delta Ra$ and $\Delta Dec$, are calculated.
The angular shifts are summarized in Table 3. The errors include only statistical errors ($1\sigma$).
The angular shifts should be caused by the motions referenced to Sgr A$^\ast$ in projection. 
The elapsed time between these observation epochs is $\Delta t=1.698$ year. As mentioned in Introduction, the Galactic center distance is assumed to be $d=8.0$ kpc.
In order to obtain the proper motion, the angular shifts are divided by the elapsed time, and times the Galactic center distance, which are given by $V_\mathrm{Ra}=d\frac{\Delta Ra}{\Delta t}$ and $V_\mathrm{Dec}=d\frac{\Delta Dec}{\Delta t}$. 
And the amplitude and position angle of the proper motion are also derived using the formulas, $V_\bot=\sqrt{V_\mathrm{Dec}^2+V_\mathrm{Ra}^2}$ and $PA=-\arctan{\frac{V_\mathrm{Ra}}{V_\mathrm{Dec}}}$. The proper motions referenced to Sgr A$^\ast$ of the detected objects are also summarized in Table 3. 
The proper motion of No.64 object is not obtained because this object was not detected at 2019.471 year. 
The proper motions by IR observations are also shown for comparison in Table 3  (\cite{Paumard}, \cite{Muzic2008}, \cite{Schodel2009}). 
The derived proper motions are mostly consistent with those by IR observations except a few cases.

\begin{figure}
\begin{center}
\includegraphics[width=15cm, bb=0 0 532 451 ]{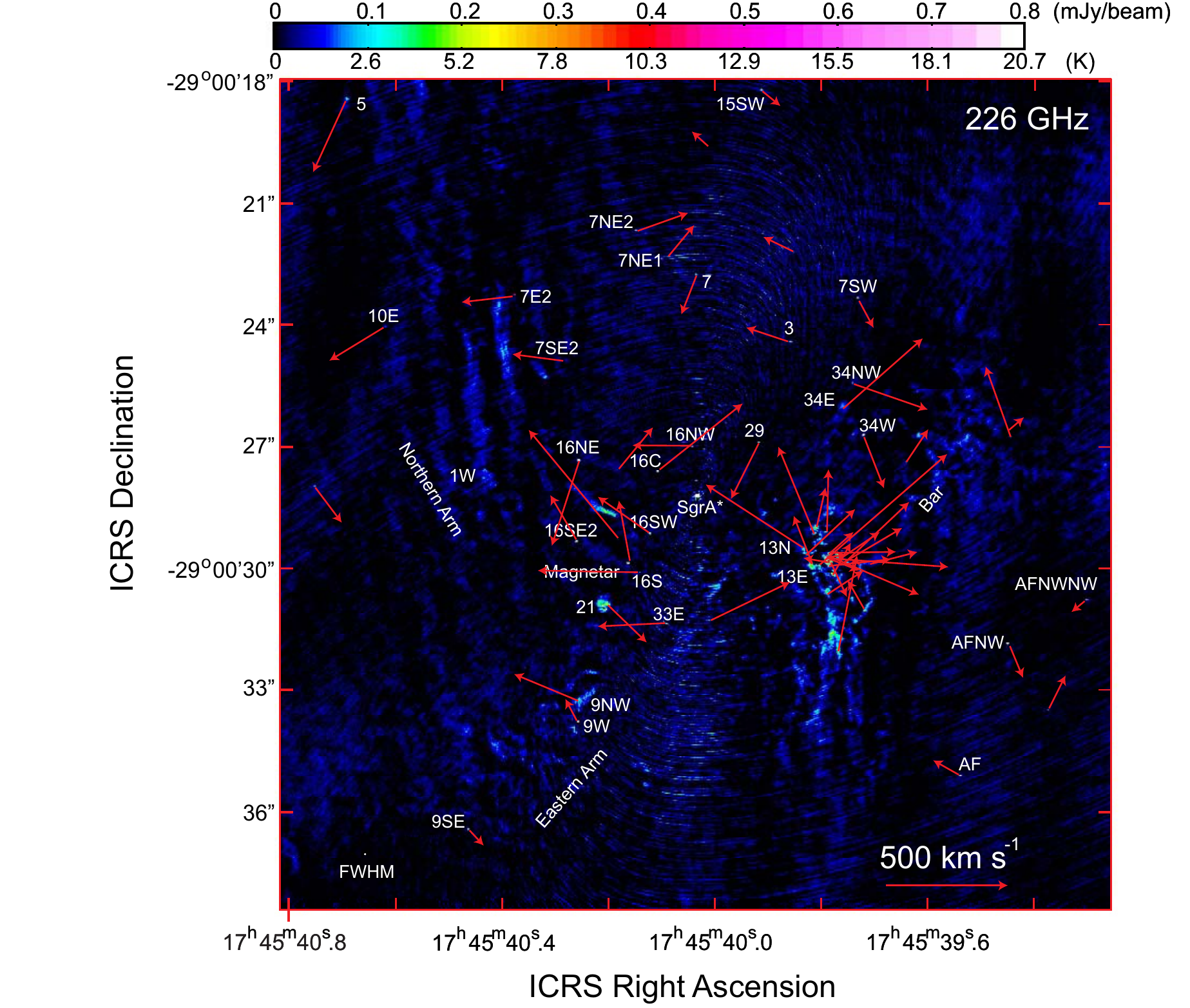}
\end{center}
 \caption{Proper motion referenced to Sgr A$^\ast$ of the detected objects, which are shown as red arrows overlaid on the continuum map at 226 GHz.
The angular resolution of the continuum map is $0\farcs029 \times 0\farcs020, PA=86^\circ$  in FWHM, which is shown as an oval at the lower left corner.  }
 \label{Fig7}
\end{figure}
\begin{figure}
\begin{center}
\includegraphics[width=10cm, bb=0 0 480 435 ]{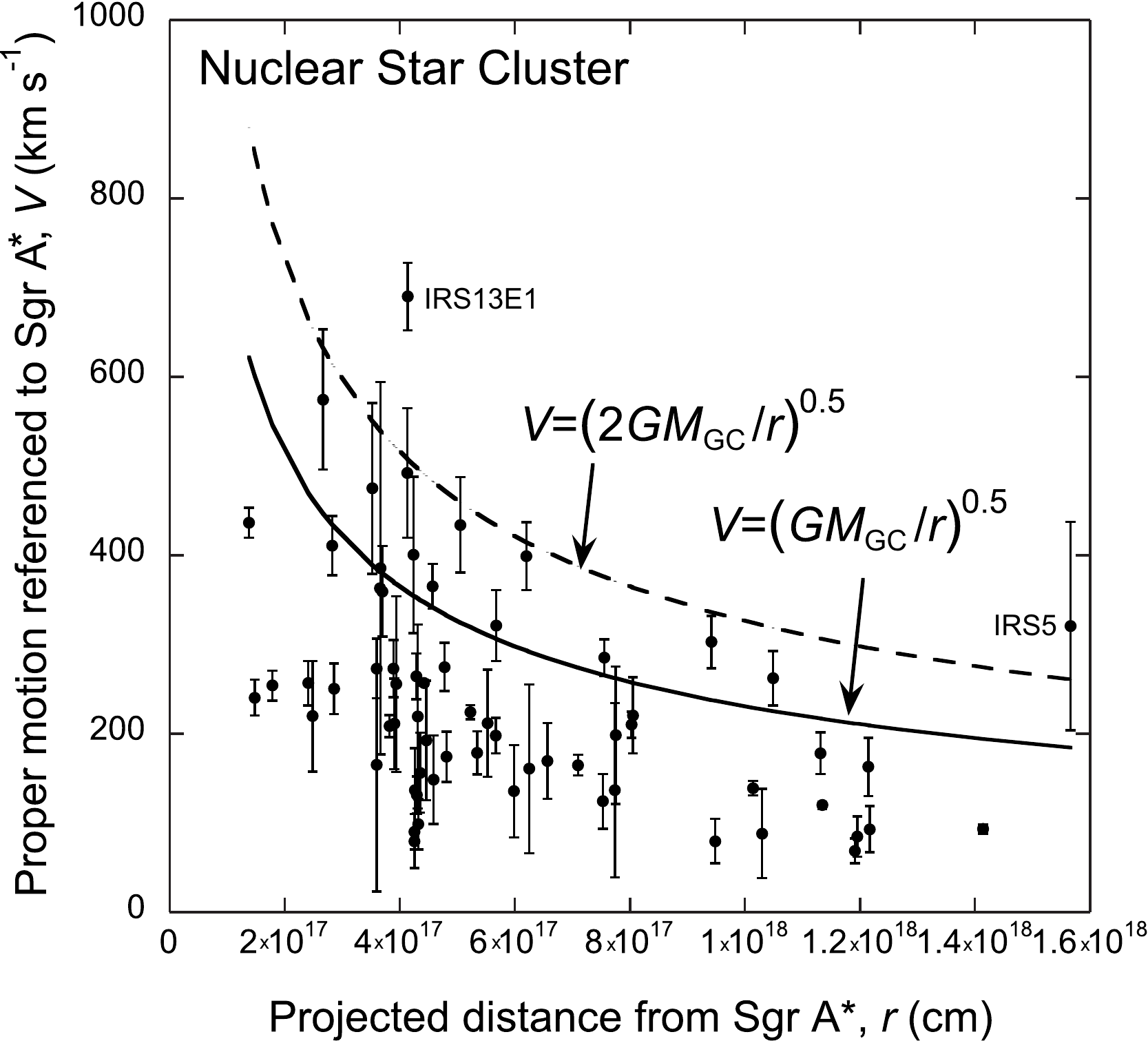}
\end{center}
 \caption{Relation between the proper motion and projected distance from Sgr A$^\ast$. The solid and broken line curves show the upper limit velocities for a circular orbit around Sgr A$^\ast$,  $V=\sqrt{\frac{GM_\mathrm{GC}}{r_\mathrm{peri}}}$, and a Kepler orbit with high eccentricity, $V=\sqrt{\frac{2GM_\mathrm{GC}}{r_\mathrm{peri}}}$, respectively.}
 \label{Fig8}
\end{figure}

Figure 7 shows the proper motions of the detected object in the NSC as arrows overlaid on the continuum map at 226 GHz  in order to explore large scale tendency of the proper motions. As mentioned above, the derived proper motions are referenced to the position of Sgr A$^\ast$. The proper motions seem to be far from random in general.  If anything, they are fairly uniform when they are in small groups, for example the IRS16, IRS13E, and IRS13N clusters. These proper motions are mostly consistent with those by IR observations (e.g. \cite{Paumard}, \cite{Muzic2008}, \cite{Schodel2009}, \cite{Yelda2014}). 
The member stars in the IRS16 cluster seem to stream to northeast in the region southeast Sgr A$^\ast$. Moreover, they seem to stream to northwest in the region  northeast Sgr A$^\ast$. Therefore,  the member stars in the IRS16 cluster including IRS29 seem to rotate around Sgr A$^\ast$ in clockwise.
The mean rotation velocity of the cluster is estimated to be $\overline{V}=343\pm149$ km s$^{-1}$.  
There may be counter clockwise motion around Sgr A$^\ast$ in the area more distant from Sgr A$^\ast$.
Because the detected objects are sparse in the outer area, it is fairly difficult to discuss the tendency. 
The proper motions in IRS13E and IRS13N clusters will be discussed separately in the following subsection.

The amplitude of the proper motions is distributed in the range from $V_\bot=69$ km s$^{-1}$ to $V_\bot=690$ km s$^{-1}$. 
Figure 8 shows the relation between the proper motion and projected distance from Sgr A$^\ast$. The solid and broken line curves in the figure show the upper limit of the rotation velocity for a circular orbit around Sgr A$^\ast$,  $V=\sqrt{\frac{GM_\mathrm{GC}}{r_\mathrm{peri}}}$, and the upper limit of the orbit velocity  of a Kepler orbit with high eccentricity when the object is at the periastron, $V=\sqrt{\frac{2GM_\mathrm{GC}}{r_\mathrm{peri}}}$, respectively. The proper motions of almost all objects are  less than the upper limit velocity for a Kepler orbit with high eccentricity. Moreover, the proper motions of many objects are less than the upper limit velocity for a circular orbit.  
This tendency is mostly consistent with those by IR observations (e.g. \cite{Paumard}, \cite{Muzic2008}, \cite{Schodel2009}, \cite{Yelda2014}).  These suggest that almost all objects are bounded by the GCBH and rotate around it.
The proper motions of IRS13E1 and IRS5 are nominally larger than the upper limit velocity for a Kepler orbit with high eccentricity. The observed proper motions of IRS13E1, $V_\mathrm{Ra}=-520\pm32$ km s$^{-1}$, $V_\mathrm{Dec}=453\pm21$ km s$^{-1}$, is not consistent with that by IR observations (e.g. $V_\mathrm{Ra}=-143\pm1$ km s$^{-1}$, $V_\mathrm{Dec}=-106\pm1$ km s$^{-1}$ in \cite{Fritz2010}, $V_\mathrm{Ra}=-195\pm8$ km s$^{-1}$, $V_\mathrm{Dec}=-102\pm5$ km s$^{-1}$ in \cite{Schodel2009}).  It is not ruled out that the observed position of IRS13E1 suffers from an unavoidable artifact as mentioned previously (see Figure 9). In addition, the fitting error of IRS5 is fairly large because the object is near the edge of the FOV. 
Therefore these deviant proper motions would be the results of these systematic errors.

\begin{figure}
\begin{center}
\includegraphics[width=14cm, bb=0 0 540 581 ]{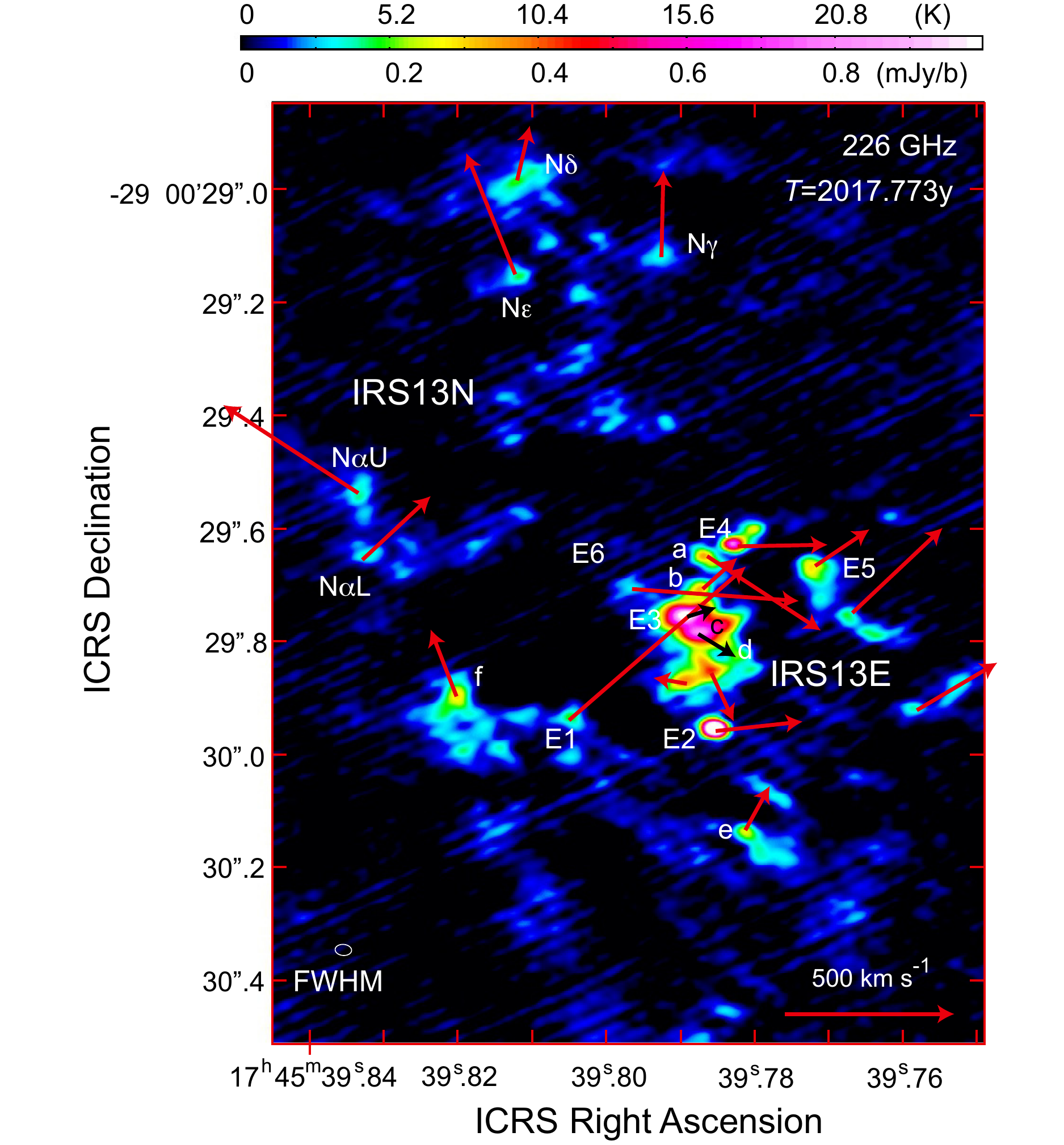}
\end{center}
 \caption{Proper motion of the IRS13E and IRS13N clusters, which are shown as red arrows overlaid on the enlarged continuum map at 226 GHz.
The angular resolution of the continuum map is $0\farcs029 \times 0\farcs020, PA=-86^\circ$  in FWHM, which is shown as an oval at the lower left corner.  The mean proper motion of the IRS13E cluster except IRS13E3 and IRS13Ec are estimated to be $\overline{V_\mathrm{Ra}}=-218\pm30$ km s$^{-1}$ and $\overline{V_\mathrm{Dec}}=73\pm30$ km s$^{-1}$. On the other hand, the mean proper motions of the IRS13N cluster are estimated to be $\overline{V_\mathrm{Ra}}=60\pm210$ km s$^{-1}$ and $\overline{V_\mathrm{Dec}}=274\pm76$ km s$^{-1}$.}
 \label{Fig9}
\end{figure}
\begin{figure}
\begin{center}
\includegraphics[width=10cm, bb=0 0 425 415 ]{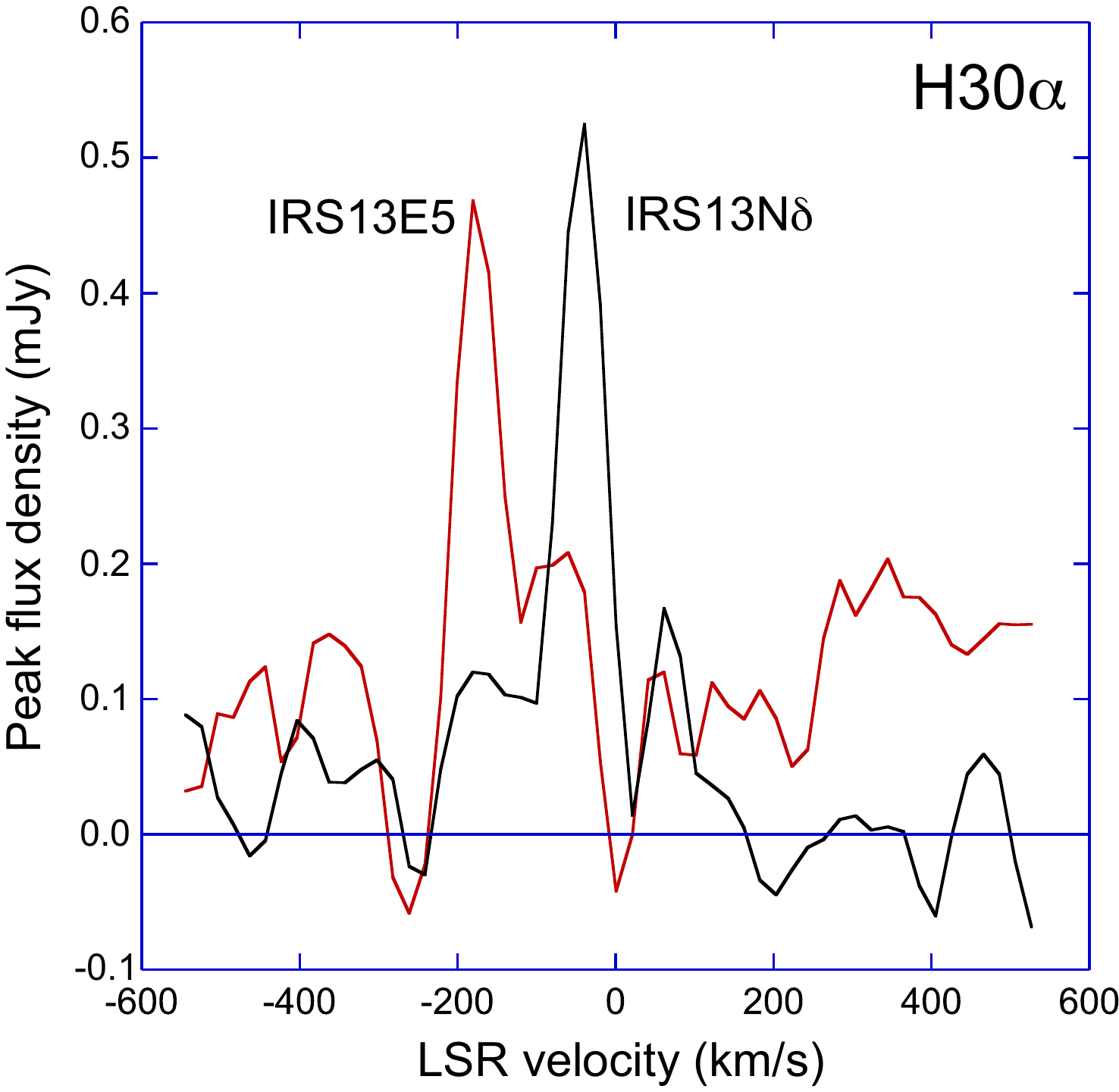}
\end{center}
 \caption{Line profiles toward IRS13E5 (in red) and IRS13N$\delta$ (in black) in the H30$\alpha$ recombination line. The sampling areas are ovals with the beam size, $0\farcs037 \times 0\farcs025$, centered at  IRS13E5 and IRS13N$\delta$ (see Table 1).}
 \label{Fig10}
\end{figure}
\subsection{Proper Motion of the Member Stars of the IRS13E and IRS13N Clusters}
Figure 9 is the enlarged map showing the proper motions of the member objects of the IRS13E and IRS13N clusters.
These clusters seem to have peculiar proper motions (see Figure 7).
Member objects of the IRS13E cluster stream monotonously to west with the velocity of  $\gtrsim200$ km s$^{-1}$. 
The peculiar proper motion, so called "co-moving cluster", has been shown by IR observations (e.g. \cite{Ott}, \cite{Maillard}, \cite{Schodel2009}).
Although IRS13E2 and IRS13E4 are observed as WR stars certainly, IRS13E3 and IRS13Ec are thought to be gas blobs not real stars based on IR spectroscopy (e.g. \cite{Fritz2010}). The proper motions of IRS13E2 and IRS13E4 are derived to be $V_\mathrm{Ra}=-256\pm3$ km s$^{-1}$, $V_\mathrm{Dec}=27\pm2$ km s$^{-1}$ and $V_\mathrm{Ra}=-265\pm23$ km s$^{-1}$, $V_\mathrm{Dec}=4\pm14$ km s$^{-1}$, respectively. 
On the other hand, the values based on IR observations are $V_\mathrm{Ra}=-276\pm6$ km s$^{-1}$, $V_\mathrm{Dec}=56\pm8$ km s$^{-1}$ and $V_\mathrm{Ra}=-254\pm7$ km s$^{-1}$, $V_\mathrm{Dec}=15\pm11$ km s$^{-1}$, respectively  (\cite{Schodel2009}). Our derived values are nearly equal to the IR values. 
The ensemble mean proper motion of the member objects except IRS13E3 and IRS13Ec are estimated to be $\overline{V_\mathrm{Ra}}=-218\pm30$ km s$^{-1}$ and $\overline{V_\mathrm{Dec}}=73\pm30$ km s$^{-1}$, respectively. These mean proper motions are mostly consistent with those by previous IR observations (e.g. \cite{Ott}, \cite{Maillard}, \cite{Schodel2009}).
Although IRS13E1, IRS13E2, IRS13E4, and IRS13E6 are not detected in the H$30\alpha$ recombination line, IRS13E3, IRS13Ec and IRS13E5 are detected in the line (see \cite{Tsuboi2019b}). Figure 10 shows the line profile toward IRS13E5. Using Gaussian fit, the central velocity and FWHM velocity width of IRS13E5 are estimated to be $V_\mathrm{C}=-169.8\pm4.9$ km s$^{-1}$ and  $V_\mathrm{FWHM}=\sqrt{\Delta V_\mathrm{obs}^2-(2sampling)^2}=67\pm12$ km s$^{-1}$, respectively. The FWHM velocity width is fairly wider than that of a hyper compact HII region even in the Galactic center region (e.g. \cite{Tsuboi2019b}).

The proper motions of IRS13E3 and IRS13Ec are fairly different from those of other member objects of the IRS13E cluster.
The  proper motions of IRS13E3 and IRS13Ec are derived to be $V_\mathrm{Ra}=-79\pm27$ km s$^{-1}$, $V_\mathrm{Dec}=10\pm13$ km s$^{-1}$ and  $V_\mathrm{Ra}=-123\pm18$ km s$^{-1}$, $V_\mathrm{Dec}=-48\pm11$ km s$^{-1}$, respectively. 
The derived proper motions of IRS13E3 are nearly equal to those by IR observations (e.g. $V_\mathrm{Ra}=-105\pm23$ km s$^{-1}$, $V_\mathrm{Dec}=-16\pm27$ km s$^{-1}$ in \cite{Schodel2009}, $V_\mathrm{Ra}=-82\pm9$ km s$^{-1}$, $V_\mathrm{Dec}=8\pm9$ km s$^{-1}$ in \cite{Fritz2010}, $V_\mathrm{Ra}=-97\pm18$ km s$^{-1}$, $V_\mathrm{Dec}=-14\pm15$ km s$^{-1}$in \cite{Eckart2013}). 
However, the proper motion of IRS13Ec is not derived by IR observations because the IR counterpart of IRS13Ec is not clearly detected (e.g. see Figure 2 in \cite{Fritz2010}).
Although IRS13E2 and IRS13E4 are located southeast and north of IRS13E3 in the IR map observed over 10 years ago (e.g. see Figure 2 in \cite{Fritz2010}),  they are located southwest and northwest of IRS13E3 in  2017 map as shown in Figure 9.
The similar positional change is also seen in the relation between IRS13E3 and IRS13E5.
These are consistent with that the proper motion of IRS13E3 is fairly smaller than those of IRS13E2, IRS13E4 and IRS13E5.
The dissimilarity of the proper motions of IRS13E3 and IRS13Ec from the rest of IRS13 cluster member objects suggests that they are not member objects of the cluster. Although the radial velocity analysis of the ionized gas in IRS13E3 has been published (\cite{Tsuboi2017b}, \cite{Tsuboi2019}), the joint analysis of the radial velocity and proper motion will be discussed in another upcoming paper.

The member objects of the IRS13N cluster have been identified in radio (e.g. \cite{Yusef-Zadeh2014}) and IR observations (e.g. \cite{Muzic2008}, \cite{Eckart2013}).
The objects of the IRS13N cluster also seem to stream monotonously to north with the velocity of $\gtrsim200$ km s$^{-1}$ as shown in Figure 9 and Table 3.  The co-moving property toward north of the IRS13N cluster had been found in IR observations (e.g. \cite{Muzic2008}).
The ensemble mean proper motions of the member stars are estimated to be $\overline{V_\mathrm{Ra}}=60\pm210$ km s$^{-1}$ and $\overline{V_\mathrm{Dec}}=274\pm76$ km s$^{-1}$. 
 IRS13N$\delta$ is also detected  as a single peak line profile object in the H$30\alpha$ recombination line (see Figure 9). Other member objects shown in Table 3 is not detected in the H$30\alpha$ recombination line. Using Gaussian fit, the central velocity and FWHM velocity width of IRS13N$\delta$ are estimated to be $V_\mathrm{C}=-43.1\pm2.3$ km s$^{-1}$ and  $V_\mathrm{FWHM}=\sqrt{\Delta V_\mathrm{obs}^2-(2sampling)^2}=59\pm16$ km s$^{-1}$, respectively.
As in the case of IRS13E5, the FWHM velocity width of IRS13N$\delta$ is also wider than that of a hyper compact HII region (e.g. \cite{Tsuboi2019b}).

\section{Conclusions}
We have observed  Sgr A$^\ast$  and the surrounding area including the NSC at 230 GHz  with ALMA  in October 2017 and analyzed the data. 
We also analyzed the similar ALMA archive data obtained in June 2019.
The angular resolution in the both resultant maps is $\sim0\farcs03$. 
We determined the positions relative to Sgr A$^\ast$ of 65 compact objects on the resultant map of October 2017 using 2-D Gaussian fit.
The positional accuracy is as high as $\sim0\farcs001$. 
We also determined the relative positions of 64 objects on the resultant map of June 2019.
We derived the proper motions relative to Sgr A$^\ast$ of the 64 objects by comparing these relative positions. 
The derived proper motions are roughly described with both clockwise and counterclockwise  rotations around Sgr A$^\ast$.  The rotation velocities are reproduced by Kepler orbits bounded around Sgr A$^\ast$.
Moreover, the proper motions include co-moving clusters for example IRS13E and IRS13N.
These positions and proper motions are almost consistent with those by previous infrared observations.
Therefore the demonstrations prove that ALMA is a powerful tool for precision astrometry of this region. 
\begin{ack} 
This work is supported in part by the Grant-in-Aid from the Ministry of Education, Sports, Science and Technology (MEXT) of Japan, No.19K03939. The National Radio Astronomy Observatory (NRAO) is a facility of the National Science Foundation operated under cooperative agreement by Associated Universities, Inc. USA.  ALMA is a partnership of ESO (representing its member states), NSF (USA) and NINS (Japan), together with NRC (Canada), NSC and ASIAA (Taiwan), and KASI (Republic of Korea), in cooperation with the Republic of Chile. The Joint ALMA Observatory (JAO) is operated by ESO, AUI/NRAO and NAOJ. This paper makes use of the following ALMA data: ADS/JAO.ALMA\#2015.1.01080.S,  ALMA\#2017.1.00503.S, and \#2018.1.001124.S.  

\end{ack}

\end{document}